\documentclass[iop,apj,tighten,twocolappendix,numberedappendix]{emulateapj}
\usepackage{apjfonts}

\usepackage{graphicx}
\usepackage{amsmath}
\usepackage{amssymb}
\usepackage{color}
\usepackage{mathtools}

\usepackage[breaklinks,colorlinks,citecolor=blue,linkcolor=red]{hyperref} 
\usepackage[all]{hypcap}

\begin{document}

\title{Gravitational waves and Intermediate Mass Black Hole retention in Globular Clusters}

\author{Giacomo Fragione\altaffilmark{1}, Idan Ginsburg\altaffilmark{2} and Bence Kocsis\altaffilmark{3}}
 \affil{$^1$Racah Institute for Physics, The Hebrew University, Jerusalem 91904, Israel} 
  \affil{$^2$Astronomy Department, Harvard University, 60 Garden St, Cambridge, MA 02138, USA}
  \affil{$^3$Institute of Physics, E\"{o}tv\"{o}s University, P\'azm\'{a}ny P. s. 1/A, Budapest, 1117, Hungary}


\begin{abstract} 
The recent discovery of gravitational waves (GW) has opened new horizons for physics. Current and upcoming missions, such as LIGO, VIRGO, KAGRA, and \textit{LISA}, promise to shed light on black holes of every size from stellar mass (SBH) sizes up to supermassive black holes. The intermediate mass black hole (IMBH) family has not been detected beyond any reasonable doubt. Recent analyses suggest observational evidence for the presence of IMBHs in the centers of two Galactic globular clusters. In this paper, we investigate the possibility that globular clusters were born with a central IMBH, which undergo repeated merger events with SBHs in the cluster core. By means of a semi-analytical method, we follow the evolution of the primordial cluster population in the galactic potential and the mergers of the binary IMBH-SBH systems. Our models predict $\approx 1000$ IMBHs within $1$ kpc from the galactic center and show that the IMBH-SBH merger rate density changes from $\mathcal{R}\approx 1000$ Gpc$^{-3}$ yr$^{-1}$ beyond $z\approx 2$ to $\mathcal{R}\approx 1-10$ Gpc$^{-3}$ yr$^{-1}$  at $z\approx 0$. The rates at low redshifts may be significantly higher if young massive star clusters host IMBHs. The merger rates are dominated by IMBHs with masses between $10^3$ and $10^4\,\mathrm{M}_{\odot}$. Currently there are no LIGO/VIRGO upper limits for GW sources in this mass range, but our results show that at design sensitivity these instruments  will detect IMBH-SBH mergers in the coming years. \textit{LISA} and the Einstein Telescope will be best suited to detect these events. The inspirals of IMBH-SBH systems may also generate an unresolved GW background.
\end{abstract}

\keywords{Galaxy: kinematics and dynamics -- stars: kinematics and dynamics -- galaxies: star clusters: general -- stars: black holes}

\section{Introduction}

Unlike supermassive (SMBHs, $M\gtrsim 10^5\ \mathrm{M}_{\odot}$) and stellar-mass black holes (SBHs, $10\ \mathrm{M}_{\odot}\lesssim M\lesssim 100\ \mathrm{M}_{\odot}$), the existence of intermediate-mass black holes (IMBHs), with masses $100\ \mathrm{M}_{\odot}\lesssim M\lesssim 10^5\ \mathrm{M}_{\odot}$, has not yet been confirmed. Assuming that the correlation between SMBHs and their stellar environments holds also for the range of IMBH masses \citep{mer01,krl13,lue13,mer13}, such compact objects may be hosted by globular clusters (GCs). Recently, \citet{bau17} suggested that $\omega$ Cen should host a $\approx 40,000\ \mathrm{M}_{\odot}$ IMBH in its center, while \citet{kiz17} showed indirect evidence of a $\approx 2,200\ \mathrm{M}_{\odot}$ IMBH in $47$ Tuc. An IMBH would remain dark if not emitting due to accretion. Some pointlike ultra-luminous X-ray (ULX) sources ($10^{39}\lesssim L_X/\mathrm{erg\ s}^{-1}\lesssim 10^{41}$) in nearby galaxies are seen to be brighter than typical accreting SBHs, but less luminous than typical active galactic nuclei, and could be explained by IMBHs \citep{fab06,dav11}. In some of these cases, the empirical mass scaling relations of quasi-periodic oscillations (QPOs) also hints at IMBH masses \citep{abr04,pas14,kaa17}. If present, an IMBH gravitationally interacts with the host cluster and influences the evolution of the GC composition \citep{bau17}. However, no direct compelling dynamical evidence of IMBHs has been found to date. In future observations, the IMBHs' dynamical effects and/or their nHz-frequency gravitational waves may be indirectly detected in the Milky Way's central nuclear star cluster \citep{gua09,gua10,koc12,mer13}. Further, IMBHs may also be detected indirectly in the Milky Way's GCs. The innermost central regions of GCs have not been resolved with sufficient precision to detect IMBHs. The accurate modeling of how IMBHs modify the stellar composition of GCs would facilitate GC-target selection for upcoming surveys \citep{luk13,lkg13,mez17}.

One avenue for the formation of IMBHs requires very dense environments as in the centers of GCs. \citet{por02} found that star clusters with small initial half-mass relaxation times are dominated by stellar collisions driven by the segregation of the most massive stars to the cluster center, where they form hard binaries. The majority of collisions occur with the same star, resulting in the runaway growth of a supermassive object, which may grow up to $0.1$\% of the mass of the entire star cluster. Similarly, \citet{fre06} showed that in less than the main sequence lifetime of massive stars ($\lesssim 3$ Myr), a single very massive star can grow up to $\approx 400$-$4000\ \mathrm{M}_{\odot}$) through a runaway sequence of mergers. \citet{mil02b} suggest that IMBHs can form from repeated mergers of a $\approx 50\ \mathrm{M}_{\odot}$ BH with other SBHs (of lower mass) in the center of a GC. Recently, \citet{gie15} found slow and fast regimes of IMBH mass growth. The larger the initial cluster concentration, the larger the probability of forming an IMBH, which forms earlier and faster.
IMBHs may also be produced in the early universe by the direct collapse of massive Pop III stars \citep{mad01,wha12,woo17}, by fragmentation of supermassive black hole (SMBH) accretion disks \citep{McKernan+2012,McKernan+2014}, or by super-Eddington accretion of stellar mass BHs in SMBH accretion disks \citep{koc11}. Further, the Milky Way hosts a number of very dense young star clusters in the inner few parsecs, which might form IMBHs as described above \citep{mer13}. The cluster may spiral into the nucleus, carrying the IMBH toward the SMBH until two black holes form a binary system \citep{fra17,pet17}, also generating bursts of hypervelocity stars \citep{cap15,fra16,fgg17,fgu18}. However, there is currently no compelling evidence of the presence of an IMBH in the galactic center \citep{mer13}, where a cusp of stars is observed \citep{frs18}. 

Several attempts have been done in modelling GCs with IMBHs. \citet{bau17} ran a large number of N-body simulations with different mass fractions $M_{\mathrm{BH}}/M_{\mathrm{GC}}$ of the central IMBH, and found indirect evidence that $\omega$ Cen could host an IMBH in its center. \citet{lut13} performed $N$-body simulations of GCs hosting IMBHs, paying attention to the SBH retention fraction and the primordial stellar binary fraction. They found that a cluster with a central IMBH usually has a shorter lifetime as a consequence of the enhanced ejection of stars due to lower mass segregation. \citet{mac16} investigated the features of the stars and compact remnants bound tightly to the IMBH across the cluster lifetime. \citet{lei14} studied the co-existence of SBH binaries in GCs with a central IMBH. Moreover, SBHs formed from the most massive progenitors could	have been embedded in a gas-rich environment for many Myrs, which can affect the BH mass distribution and dynamics \citep{lee13}.

Gravitational wave (GW) astronomy will help in the hunt for the first direct evidence of IMBHs. IMBH-SBH binaries may form in GCs and represent a down-scaled version of extreme mass-ratio inspirals, the inspiral of a stellar BH into a SMBH \citep{hop06,ama07}. LIGO\footnote{\url{http://www.ligo.org}}, Einstein Telescope\footnote{\url{http://www.et-gw.eu}} (ET), and \textit{LISA}\footnote{\url{https://lisa.nasa.gov}} will be able to detect IMBH-SBH binaries of different total mass (small, intermediate and massive, respectively, \citealt{ama10,aep10}). After an IMBH-BH binary forms in a GC, it may produce an IMBH-SBH merger. \citet{man08} have called such events intermediate mass-ratio inspirals (IMRIs). When BHs merge as a consequence of GW emission, the merger product will be imparted a gravitational wave recoil kick which, depending on the symmetric mass-ratio $\eta=q/(1+q)^2$ and spins (here $q$ is the BH mass ratio) of the individual BHs, may be up to several thousands km s$^{-1}$ times $\eta^2$ \citep{lou11}. Due to the small $\eta$ of an IMRI, such recoils may be in many cases not large enough to expel the IMBH from the cluster. However, \citet{hol08} showed that there is a significant problem in retaining low-mass IMBHs in the merger-rich environment of GCs. They computed that only $3$ of the Milky Way's GCs can retain an IMBH with mass of $200\ \mathrm{M}_{\odot}$, while around $60$ GCs would retain IMBHs with an initial mass of $1000\ \mathrm{M}_{\odot}$. \citet{kon13} found an IMRI in one of their simulations, which lead to a merger with a recoil velocity higher than the escape velocity of their simulated GC. Thus, besides forming an IMBH, the challenge is also to retain it inside a star cluster as a consequence of the repeated gravitational wave kicks following mergers of IMBH-SBH. 

In this paper, we address the question whether the primordial population of GCs that formed in Milky-Way like galaxies can retain their IMBH and examine the expected distribution of such IMBHs within their host galaxies. In the Milky Way, a few percent of the stars were born in $\approx 7000$ primordial GCs \citep{gne14}. Even if a small fraction of such GCs hosted IMBHs, the number of IMRIs may be important for GW astronomy and could shed light on the origin and evolution of IMBHs. To do that, we model the evolution of GCs in the Galactic field along with the dynamics of the sub-cluster of IMBH and SBHs which form in the cluster core. We use a semi-analytical method to follow the evolution of GCs within the host galaxy \citep{gne14} and use a Monte Carlo method to follow the evolution of the IMBH within the globular cluster. We account for binary formation via dynamical interactions with stellar encounters, gravitational wave emission, general relativistic recoil kicks, and track the evolution of the spin magnitude and direction. While simulations of star clusters with IMBH-SBH coalescences have generally excessive computational costs \citep{gul04,kon13,lei14}, our semi-analytical method allows to self-consistently model and study the dynamical evolution of the sub-cluster of IMBH and SBHs embedded in thousands of primordial GCs, while they lose mass and sink towards the galactic center on cosmic timescales.

The paper is organized as follows. In Section \ref{sect:gcev}, we present the semi-analytical method we use to evolve the primordial GC population on cosmic timescales. In Section \ref{sect:imbh}, we analyze the typical dynamics of the sub-cluster of IMBH and SBHs. We describe our numerical setup in Section \ref{sect:numerical} and present our results in Section \ref{sect:res}. In Section 6, we present our predictions for the rate of IMBH-SBH merger events. Finally, in Section \ref{sect:conclusions} we draw our conclusions.

\section{Globular Cluster evolution}
\label{sect:gcev}

We discuss the equations used for evolving the GC population \citep[for details see][and references therein]{gne14}. We assume that the cluster formation rate was a fixed fraction $f_{\mathrm{GC},i}$ of the overall star formation rate
\begin{equation}
\frac{dM_{\mathrm{GC}}}{dt}=f_{\mathrm{GC},i}\frac{dM_{*}}{dt}\ .
\end{equation}
We set $f_{\mathrm{GC},i}=0.011$. We assume that clusters formed at redshift $z=3$ and calculate their evolution for $11.5$ Gyr until today \citep{gne14}. The initial mass of the clusters is drawn from a power-law distribution
\begin{equation}
\frac{dN_{\mathrm{GC}}}{dM_{\mathrm{GC}}}\propto M_{\mathrm{GC}}^{-\beta},\ \ \ \ M_{\min}<M<M_{\max}\ .
\label{eqn:gcmassini}
\end{equation}
We adopt $\beta=2$, $M_{\min}=10^4\ \mathrm{M}_{\odot}$ and $M_{\max}=10^7\ \mathrm{M}_{\odot}$. Results do not depend on the choice of $M_{\min}$ since light clusters are expected to be disrupted by the Galactic tidal field, while only slightly on the choice of $M_{\max}$ \citep{gne14}.

After formation, GCs loose mass via three mechanisms, i.e. stellar winds, dynamical ejection of stars through two-body relaxation, and removing stars by the galactic tidal field \footnote{The last two mechanisms are not completely independent, since it is two-body relaxation that pushes stars across the tidal boundary. Anyways, when the GC goes deep in the galactic potential, the stripping will be the dominant process.}. Following \citet{pri08} and \citet{gne14}, we model the stellar mass loss considering \citet{kro01} initial mass function and using the main-sequence lifetimes from \citet{hur00}, the stellar remnant masses from \citet{che90}. We take into account the mass loss due to two-body relaxation and stripping by the galactic tidal field according to
\begin{equation}
\frac{dM}{dt}=-\frac{M}{\min(t_{\mathrm{tid}},t_{\mathrm{iso}})}\ ,
\end{equation}
where
\begin{equation}
t_{\mathrm{tid}}(r,M)\approx 10 \left(\frac{M}{10^5\ \mathrm{M}_{\odot}}\right)^{\alpha}P(r)\ \mathrm{Gyr}
\end{equation}
is the typical tidal disruption time \citep{gie08}, and
\begin{equation}
P(r)=41.4\left(\frac{r}{\mathrm{kpc}}\right)\left(\frac{V_{\mathrm{c}}(r)}{\mathrm{km}\ \mathrm{s}^{-1}}\right)^{-1}
\end{equation}
is the normalized rotational period of the cluster orbit, which takes into account the strength of the local Galactic field, and $V_{\mathrm{c}}(r)$ is the circular velocity at a distance $r$ from the galactic center. We assume $\alpha=2/3$ \citep{gie08,gne14}. In case of strong tidal field ($t_{\mathrm{tid}}<t_{\mathrm{iso}}$) the loss of stars is dominated by the Galactic tidal stripping, while in the limit of a weak tidal field ($t_{\mathrm{tid}}>t_{\mathrm{iso}}$), the evaporation of stars is controlled by internal dynamics. We compute the evaporation time in isolation as a multiple of the half-mass relaxation time \citep{gie11,gne14}
\begin{equation}
t_{\mathrm{iso}}(M)\approx 17 \left(\frac{M}{2\times 10^5\ \mathrm{M}_{\odot}}\right)\ \mathrm{Gyr}\ .
\end{equation}
When a cluster arrives near the galactic center, the tidal forces may be strong enough to tear the cluster apart, which happens when the stellar density at a characteristic place in the cluster, such as the core or half-mass radius, falls below the mean ambient density \citep{ant13}. Following \citet{gne14}, we adopt the average density at the half-mass radius
\begin{equation}
\rho_{\mathrm{h}}=10^3\frac{\mathrm{M}_{\odot}}{\mathrm{pc}^3}\min\left\{10^2,\max\left[1,\left(\frac{M}{2\times 10^5\ \mathrm{M}_{\odot}}\right)^2\right]\right\}\ .
\label{eqn:rhalfm}
\end{equation}
This equation limits $\rho_{\mathrm{h}}$ to $10^5\ \mathrm{M}_{\odot}$ pc$^{-3}$ in the most massive clusters, that is about the highest observed half-mass density. A cluster is considered disrupted if the average density at the half-mass radius is smaller than the mean ambient density
\begin{equation}
\rho_{\mathrm{h}}<\rho_*(r)=\frac{V_{\mathrm{c}}^2(r)}{2\pi G r^2}\ ,
\label{eqn:dens}
\end{equation}
due to the adopted field stellar mass, as well as the growing mass of the nuclear stellar cluster. As the nuclear cluster begins to build up, its stellar density will exceed even the high density of infalling GCs and these clusters will be directly disrupted before reaching the galaxy center \citep{gne14}.

We consider the cluster moving on a circular trajectory of radius $r$ and set this radius to be the time-averaged radius of the true, likely eccentric, cluster orbit \citep{gne14}. We consider the effect of dynamical friction on cluster orbits by evolving the radius $r$ of the orbit according to \citep{bin08}
\begin{equation}
\frac{dr^2}{dt}=-\frac{r^2}{t_{\mathrm{df}}}\ ,
\label{eqn:dynf}
\end{equation}
where
\begin{equation}
t_{\mathrm{df}}(r,M)\approx 0.45 \left(\frac{M}{10^5\ \mathrm{M}_{\odot}}\right)^{-1}\left(\frac{r}{\mathrm{kpc}}\right)^2\left(\frac{V_{\mathrm{c}}(r)}{\mathrm{km}\ \mathrm{s}^{-1}}\right)\ \mathrm{Gyr}\ .
\end{equation}
Several authors have shown the importance of the details of the cluster orbit and its relation with the local tidal field \citep{tio16,mad17}. \citet{web14} have shown that eccentric orbits increase the mass-loss rate and cluster velocity dispersion and shorten the GC relaxation time. Since the initial distribution of globular cluster eccentricities are not well understood, and to keep things simple, we include the effect of the deviation of the GC orbit from circular by including an eccentricity correction factor $f_e=0.5$ in the dynamical friction equations, consistent with the results of simulations by \citet{jia08}.

We describe the Milky Way's potential following \citet{gne14} with a central black hole ($M_{\mathrm{BH}}=4 \times 10^6$ M$_{\odot}$), a Sersic profile \citep{ser63} (see also \citealt{ter05}) with total mass $M_{\mathrm{S}}=5\times 10^{10}$ M$_{\odot}$ and effective radius $R_{\mathrm{e}}=4$ kpc, and a dark matter halo ($M_{\mathrm{DM}}=10^{12}$ M$_{\odot}$ and $r_{\mathrm{s}}=20$ kpc, \citealt{nfw97}) (see also \citealt{fao17} and \citealt{frl17}). We continuously update the Galactic mass distribution to include the gaseous and stellar debris from the disrupted GCs as the nuclear star cluster begins to form \citep{gne14}.

\section{IMBH merger history and evolution}
\label{sect:imbh}

Because of the short lifetimes of massive stars, they collapse into SBHs soon after the cluster formation. Assuming a standard \citet{kro01} initial mass function, the number of SBHs is roughly proportional to the initial cluster mass \citep{ole06}
\begin{equation}
N_{\mathrm{SBH}}\approx 3\times 10^{-3} \frac{M_{\mathrm{GC}}}{\mathrm{M}_{\odot}}
\label{eqn:nsbh}
\end{equation}
Due to dynamical friction, the SBHs segregate towards the GC center on timescales
\begin{equation}
t_{\mathrm{seg}}\approx \frac{\bar{m}}{M_{\mathrm{SBH}}} t_{\mathrm{rh}}\ ,
\end{equation}
where $\bar{m}$ and $M_{\mathrm{SBH}}$ are the average stellar mass and SBH mass, respectively, and $t_{\mathrm{rh}}$ is the half-mass relaxation time. In a few tens of Myrs, the SBHs form a self-gravitating subsystem within the core, dynamically decoupled from the rest of the cluster, until recoupling with the rest of the stars \citep{spi69}. The dynamical evolution of the SBH subsystem proceeds on a timescale shorter than the cluster dynamical timescale by a factor $\approx N_{\mathrm{BH}}/N$. As widely discussed in the literature, these SBHs can undergo strong gravitational interactions with other BHs and are likely ejected from the cluster, causing a progressive depletion of the BH sub-cluster. If a GC is Spitzer unstable, very high central SBH densities are reached and the central IMBH can grow as due to subsequent mergers with SBHs \citep{mil02b}. While Spitzer instability criterion is well satisfied from the beginning of the clusters evolution, SBHs may recouple to the rest of the cluster at later times as the cluster evolves and loses mass. For instance, for a GC mass of $10^5\ \mathrm{M}_{\odot}$, the Spitzer criterion would require less than $\approx 30$-$50$ SBH to be stable, assuming that all the SBHs are $10\ \mathrm{M}_{\odot}$. SBH mass function would even reduce this number \citep{ole06}. Recently, \citet{mor13,mor15} have used more realistic Monte-Carlo simulations of globular clusters with a BH mass spectrum. Their results have shown that many of the old massive globular clusters could still retain tens of SBHs up to present day, making such clusters still Spitzer unstable \citep[see also][]{bre13,arc16}.

In case the center of the cluster hosts an IMBH, one of the SBHs forms a bound pair with the central IMBH. \citet{lei14} showed that such an event takes $\lesssim 100$ Myr of cluster evolution. Then, the remaining SBHs undergo strong interactions with the central SBH-IMBH binary, which may merge and be ejected out of the cluster because of GW recoil kick. If not, the IMBH merger remnant will capture another SBH on a timescale comparable to the timescale of SBH ejection \citep{lei14}. Typically, the second SBH companion is lighter then the first one and $T_{\mathrm{GW}}$ becomes longer, decreasing the rate of IMBH-SBH mergers. \citet{gul04} found that IMBH-SBH binary continuously interacts with objects in the cluster and will have a very high eccentricity after its last encounter before inspiraling and merging due to GWs, with a large fraction retaining a measurable eccentricity ($0.1\lesssim e\lesssim 0.2$). \citet{ole06} found that the typical merger rate is made up of two phases, the first when the cluster is undergoing a lot of binary interactions and the second when the binary fraction is depleted and nearly zero, which scales roughly as $t^{-1}$. The typical timescale for the central dark sub-cluster to deplete all the SBHs may range from a few $100$ Myr to several Gyr depending on the cluster mass, more massive being longer. Actually, the timescale for all BHs to be ejected increases with larger cluster mass both because the initial number of SBHs is larger and because the relaxation time is longer \citep{lei14}. In the mean time, the host GC loses mass and shrinks its orbit as a consequence of dynamical friction (Section \ref{sect:gcev}).

During the first moments of its lifetime, the binary SBH-IMBH has usually very high eccentricities and typically decreases its semi-major axis due to dynamical friction and later to scattering slingshots with ambient objects at the typical hardening radius \citep{mim01,mer13}
\begin{equation}
a_{\mathrm{h}}=\frac{M_{\mathrm{SBH}}}{M_{\mathrm{IMBH}}+M_{\mathrm{SBH}}} \frac{r_{\mathrm{inf}}}{4}\ ,
\end{equation} 
where $r_{\mathrm{inf}}=GM_{\mathrm{IMBH}}/\sigma_c^2$ is the influence radius of the IMBH, $\sigma_c$ the cluster central velocity dispersion, $M_{\mathrm{IMBH}}$ and $M_{\mathrm{SBH}}$ are the masses of the IMBH and SBH, respectively, and $a_{\mathrm{h}}$ ranges from a few AU in the most massive clusters to a few $100$ AU in the lightest ones, that have smaller velocity dispersions and less massive IMBHs. The binary is hardened by three and four-body interactions, and may form stable hierarchical triples where the Kozai-Lidov resonances become important \citep{koz62,lid62} on a typical timescale
\begin{equation}
T_{LK}\approx \frac{P}{2\pi}\frac{M_{\mathrm{IMBH}}+M_{\mathrm{SBH}}}{M_{\mathrm{SBH},\mathrm{out}}}\left(\frac{a_{\mathrm{out}}}{a_{\mathrm{in}}}\right)^3 (1-e_{\mathrm{out}}^2)^{3/2}\ .
\end{equation}
Her, $P$ is the period of the IMBH-SBH binary, $M_{\mathrm{SBH},\mathrm{out}}$ is the mass of the external companion, $a_{\mathrm{in}}$ the IMBH-SBH semi-major axis, and $a_{\mathrm{out}}$ and $e_{\mathrm{out}}$ the semi-major axis and eccentricity of the outer orbit, respectively. High eccentricities are fundamental to make the binary enter the GW regime even at high semi-major axis, and the Kozai-Lidov oscillations may play an important role in increasing the eccentricity of the IMBH-SBH binary. Three-body interactions may lead to the ejection of the binary if its semi-major axis is below \citep{ant16}
\begin{equation}
a_{\mathrm{ej}}=\frac{0.2 G\mu M_{\mathrm{SBH},\mathrm{out}}^2}{(M_{\mathrm{IMBH}}+M_{\mathrm{SBH}})(M_{\mathrm{IMBH}}+M_{\mathrm{SBH}}+M_{\mathrm{SBH},\mathrm{out}}) v_{esc}^2}\ ,
\label{eqn:aej}
\end{equation}
where $\mu$ is the binary reduced mass. As the IMBH-SBH binary hardens, the typical time to the next interaction falls below the GW timescale, when the semi-major axis of the binary becomes smaller than \citep{has16} \begin{equation}
a_{\mathrm{GW}}\approx \frac{0.06\ \mathrm{AU}}{(1-e^2)^{7/10}}\left(\frac{M_{\mathrm{IMBH}}}{10^2\ \mathrm{M}_{\odot}}\right)^{1/5}\left(\frac{10^5\ \mathrm{pc}^{-3}}{n_{\mathrm{c}}}\right)^{1/5}\ ,
\label{eqn:agw}
\end{equation}
where $e$ is the eccentricity of the IMBH-SBH binary and $n_{\mathrm{c}}$ is the number density of stars and BHs in the cluster center. If $a_{\mathrm{GW}}>a_{\mathrm{ej}}$, the merger occurs before the binary is ejected via three-body interactions. At this point the semi-major axis and eccentricity of the IMBH-BH binary evolve according to \citep{pet64}
\begin{equation}
\frac{da}{dt}=-\frac{64}{5}\frac{G^3 M_{\mathrm{IMBH}} M_{\mathrm{SBH}} M}{c^5 a^3 (1-e^2)^{7/2}}\left(1+\frac{73}{24}e^2+\frac{37}{96}e^4\right)
\label{eqn:semgw}
\end{equation}
\begin{equation}
\frac{de}{dt}=-\frac{304}{15}\frac{G^3 M_{\mathrm{IMBH}} M_{\mathrm{SBH}} M}{c^5 a^4 (1-e^2)^{5/2}}\left(e+\frac{121}{304}e^3\right)\ ,
\label{eqn:eccgw}
\end{equation}
and the binary merges within
\begin{equation}
T_{\mathrm{GW}}=\frac{3}{85}\frac{a^4 c^5}{G^3 M_{\mathrm{IMBH}} M_{\mathrm{SBH}} M}(1-e^2)^{7/2}\ ,
\end{equation}
where $M=M_{\mathrm{IMBH}}+M_{\mathrm{SBH}}$. \citet{lei14} illustrated that in N-body simulations, $T_{\mathrm{GW}}$ may be even of the order of $1$ Myr or less because of the very high eccentricities reached by the IMBH-BH binary.

As shown by $N$-body simulations \citep{kon13,has16}, the merging of the binary IMBH-SBH happens after a few Myrs. When a merging occurs, the merger remnant undergoes a recoil kick and acquires a velocity \citep{lou12}
\begin{equation}
\textbf{v}_{\mathrm{kick}}=(1+e)\left[v_m \hat{e}_{\perp,1}+v_{\perp}(\cos \xi \hat{e}_{\perp,1}+\sin \xi \hat{e}_{\perp,2})+v_{\parallel} \hat{e}_{\parallel}\right]\ ,
\label{eqn:vkick}
\end{equation}
where
\begin{equation}
v_m=A\eta^2\sqrt{1-4\eta}(1+B\eta)
\end{equation}
\begin{equation}
v_{\perp}=\frac{H\eta^2}{1+q}(\chi_{2,\parallel}-q\chi_{1,\parallel})
\end{equation}
\begin{eqnarray}
v_{\parallel}&=&\frac{16\eta^2}{1+q}[V_{1,1}+V_A \tilde{S}_{\parallel}+V_B \tilde{S}^2_{\parallel}+V_C \tilde{S}_{\parallel}^3]\times \nonumber\\
&\times & |\mathbf{\chi}_{2,\perp}-q\mathbf{\chi}_{1,\perp}| \cos(\phi_{\Delta}-\phi_{1})\ .
\end{eqnarray}
In Eq. (\ref{eqn:vkick}), we included a term $(1+e)$ to take into account the eccentricity contribution for eccentric orbits, since IMBH-SBH binaries may have not completely circularized by the time of the merger \citep{sop07,hol08}. In the previous equations, $\eta=q/(1+q)^2$ is the symmetric mass ratio. The symbols $\perp$ and $\parallel$ refer to the direction perpendicular and parallel to the orbital angular momentum, respectively, while $\hat{e}_{\perp}$ and $\hat{e}_{\parallel}$ are orthogonal unit vectors in the orbital plane. Moreover
\begin{equation}
\tilde{\mathbf{S}}=2\frac{\mathbf{\chi}_{2,\perp}+q^2\mathbf{\chi}_{1,\perp}}{(1+q)^2}
\end{equation}
and $A=1.2\times 10^4$ km s$^{-1}$, $H=6.9\times 10^3$ km s$^{-1}$, $B=-0.93$, $\xi=145^{\circ}$ \citep{gon07,lou08}, and $V_{1,1}=3678$ km s$^{-1}$, $V_A=2481$ km s$^{-1}$, $V_B=1793$ km s$^{-1}$, $V_C=1507$ km s$^{-1}$ \citep{lou12}. Finally, $\phi_{1}$ is the phase angle of the binary and $\phi_{\Delta}$ is that between the in-plane component of the vector $\Delta_{\perp}$ of the vector
\begin{equation}
\mathbf{\Delta}=M^2\frac{\mathbf{\chi}_{2}-q\mathbf{\chi}_{1}}{1+q}\ .
\end{equation}

During a merger, gravitational waves radiate not only linear momentum, but also angular momentum and energy. In our calculations, we adjust the total spin and mass of the merger remnant to account for these losses \citep{lou10}. This allows us to follow the remnant IMBH spin and mass self-consistently.

\section{Numerical setup}
\label{sect:numerical}

We evolve the primordial GC population by means of the equations described in Section \ref{sect:gcev}, according to the prescriptions in \citet{gne14}. We assume that clusters formed at redshift $z=3$ and calculate their evolution for $11.5$ Gyr until today or their eventual tidal disruption or evaporation. Different models predict different initial seeds for the IMBH \citep{hol08}. \citet{por05} showed via $N$-body simulations that runaway collisions in a dense stellar cluster give a typical mass
\begin{equation}
M_{\mathrm{IMBH}}\approx m_{\mathrm{s}}+4\times 10^{-3} f_{\mathrm{c}} \ln \Lambda\ M_{\mathrm{GC}}\ ,
\label{eqn:imbh1}
\end{equation}
where $m_{\mathrm{s}}=50\ \mathrm{M}_{\odot}$ is the mass of the seed heavy star that initiates the runaway mergers, $\ln\Lambda=10$ is the Coulomb logarithm and $f_{\mathrm{c}}=0.2$ is a runaway efficiency factor. The previous equation gives IMBH with $\approx 1$\% of the mass of the cluster. \citet{ses12} proposed to derive the IMBH masses with a low-mass extrapolation of the M-$\sigma$ relation observed in galactic bulges \citep{mer13}
\begin{equation}
M_{\mathrm{IMBH}}=2\times 10^6 \sigma_{70}^4\ \mathrm{M}_{\odot}\ ,
\end{equation}
where $\sigma_{70}$ is the velocity dispersion in units of $70$ km s$^{-1}$. This equation typically gives a mass one order of magnitude smaller than Eq. (\ref{eqn:imbh1}). IMBHs may also have been generated as a consequence of the collapse of a Pop III star. The IMBH remnant may be as massive as a few hundreds of solar masses \citep{mad01}, or several thousands of solar masses via extreme accretion flows \citep{woo17}. Such a conclusion depends significantly on the mass of the Pop III star progenitor, which should have been thousands of solar masses, and on their initial mass function, both being highly uncertain. Predictions on the IMBH mass in the accretion disk fragmentation scenario is even more uncertain \citep{McKernan+2012,McKernan+2014}. In our simulations, we considered all the clusters hosting an IMBH at their center from the beginning of the cluster evolution. This conclusion is supported by the fast scenario discussed in \citet{gie15}, as well as by Pop III star origin. On the other hand, the gradual and progressive formation and evolution of an IMBH across the cluster lifetime would reduce the predicted rate of IMRIs at high redshift. We simply generate IMBH masses by scaling the total mass of the GC. We choose the fraction of the GC mass in IMBH respectively as $f=0.5$\%, $1$\%, $2$\% and $5$\%, in order to span all the possible regimes predicted by the previous considerations. Since we assume for the cluster mass that $M_{\min}=10^4\ \mathrm{M}_{\odot}$ and $M_{\max}=10^7\ \mathrm{M}_{\odot}$, the minimum and maximum mass of the IMBHs 
is between $50\ \mathrm{M}_{\odot}$ and $\ 5\times 10^5\mathrm{M}_{\odot}$. By scaling the mass of the GCs, we implicitly adopt a similar IMBH initial mass function as used for clusters (Equation~\ref{eqn:gcmassini}), i.e. a power-law with negative index $\beta=2$.

For SBHs, we adopt a power-law distribution of masses
\begin{equation}
\frac{dN_{\mathrm{SBH}}}{dM_{\mathrm{SBH}}}\propto M_{\mathrm{SBH}}^{-\zeta},\ \ \ \ M_{\mathrm{S},\min}<M<M_{\mathrm{S},\max}\ ,
\label{eqn:mass}
\end{equation}
where $M_{\mathrm{S},\min}=5\ \mathrm{M}_{\odot}$ and $M_{\mathrm{S},\max}=40\ \mathrm{M}_{\odot}$. We study the dependence of our results as a function of the exponent of the power-law by taking $\zeta=1$, $2$, $3$, $4$ \citep{ole16}. The recoil velocity depends on the mass ratio $q$ and is maximum for $q\approx 0.4$ (see Equation (\ref{eqn:vkick}) and \citealt{hol08}).

The spin of black holes is still uncertain. The spin of SBH is mainly determined at birth and depends on the mass of the progenitor star, its spin rate and interior structure. The spin impacts both the recoil velocity and the gravitational radiation waveforms, in particular if the spins are misaligned with the orbital axis \citep{mil15}. Recently, \citet{fis17} showed that for SBHs forming from a hierarchical sequence of mergers the distribution of spin magnitudes has peak at $\chi\approx 0.7$. Heavier SBHs that form from massive stars are expected to have low spins \citep{ama16,kus16,bel17}, but their spins may become high due to dynamical effects \citep{zal17}. For IMBHs, the situation is even more complicated \citep{mil15}. In our simulations, we consider several values for the initial IMBH and SBH spins  $\chi=0$, $0.2$, $0.5$, $0.7$, respectively, and account for the effect of the spin in the GW recoil velocity. We note that the equations we use to compute the corrections to the spin of the merger product are expected to give reliable results for non-extreme intrinsic spin magnitudes $\lesssim 0.8$-$0.9$. All the relevant relative orientations between the spins, angular momenta and orbital plane are generated according to the prescriptions in \citet{lou10}.

The eccentricity of the IMBH-SBH binary plays a role in driving the binary towards merger due to GW emission. Simulations show that the binary orbital eccentricity is usually of order of unity soon after the dynamical formation, and also as a consequence of the repeated interactions with the stellar surroundings \citep{lei14}. However, as eccentricity increases, GW emission may eventually dominate the evolution of binary semi-major axis and eccentricity which leads to the circularization of the IMBH-BH binary on shorter timescales than the average timescale between encounters with the stellar surroundings \citep{kon13,lei14}. \citet{gul06} and \citet{ole06} showed that the binaries could merge with high eccentricity for GWs detected by \textit{LISA}, while the binary will have circularized by the time GWs are detected by LIGO. Equation (\ref{eqn:vkick}) depends on the eccentricity through the factor $(1+e)$, which is valid only for small eccentricities \citep{hol08}. The exact form of the kick velocity when $e$ approaches high values is not well known, but in general $v_{\mathrm{kick}}$ becomes larger with larger eccentricities. In our fiducial models, we consider circular orbits. Additionally, we run a simulation where all binaries have $e=0.2$, which makes the recoil velocity $1.2$ times larger according to Eq. (\ref{eqn:vkick}).

The other two parameters to specify are the average number of IMBH-SBH collisions $N_{\mathrm{coll}}$ and the average time $t_{\mathrm{coll}}$ between subsequent collisions. \citet{hol08} used $N_{\mathrm{coll}}=25$, derived from simulations of \citet{gul06}, assuming that the IMBH-SBH merging phase takes place in the first moments of a GC lifetime. Moreover, \citet{hol08} found that their results were quite insensitive to $N_{\mathrm{coll}}$. We note that $N_{\mathrm{coll}}$ represents the maximum number of collisions that an IMBH can undergo in a GC. If the IMBH is ejected as a consequence of GW recoil velocity or the host GC dissolves due to Galactic tidal disruption, the number of mergers in that GC will be smaller than $N_{\mathrm{coll}}$. $t_{\mathrm{coll}}$ is hard to define and find an analytical expression for all the clusters mass. As reference, \citet{ole06} found that the typical merger rate may be described by two phases, the first when the cluster is undergoing a lot of binary interactions and the second when the binary fraction is depleted and nearly zero, which scales roughly as $t^{-1}$, while \citet{lei14} found that this typical timescale for the central dark sub-cluster to deplete all the SBHs ranges from a few $100$ Myr to several Gyr depending on the cluster mass. \citet{kon13} evolved a $2\times 10^4\ \mathrm{M}_{\odot}$ GC with a central $500\ \mathrm{M}_{\odot}$ IMBH, and found that the first merger happens $\approx 50$ Myr after the beginning of the simulations. \citet{has16} found a merging event after $\approx 110$ Myr in $N$-body simulations of a $100\ \mathrm{M}_{\odot}$ IMBH embedded in a cluster with $32$ $10\ \mathrm{M}_{\odot}$ BHs and $32000$ $1\ \mathrm{M}_{\odot}$ stars. To find a simple value of $t_{\mathrm{coll}}$ for all the clusters is not straightforward. We note that $t_{coll}$ may be quantified based on N-body simulations calibrated to IMBH-SBH binaries embedded in a star cluster \citep{sig93,lei11}. To keep things simple, we adopt $t_{\mathrm{coll}}=50$\,Myr in our fiducial model, and run additional models with $t_{\mathrm{coll}}=100$, $150$, $200$ Myr, respectively, to study the dependence of the results on this uncertain parameter, in agreement with estimates by \citet{mll02}. Moreover, we set the maximum number of IMBH-SBH merger events to $N_{\mathrm{coll}}=T_{\mathrm{life}}/t_{\mathrm{coll}}$, where $T_{\mathrm{life}}$ is the maximum lifetime of GCs. In our simulations, we assume that all the clusters formed at $z=3$, which corresponds to $T_{\mathrm{life}}=11.5$ Gyr.

In our simulations, we only consider the formation and evolution via GW emission of IMBH-BH binaries embedded in the innermost BH sub-cluster. Moreover, we do not track the dynamical encounters of the IMBH-SBH binary with other SBHs that may kick the binaries out of the host GC and simply assume that each of them merges at fixed time intervals due to the combined effect of these interactions and GWs, independently of their mass. Finally, we note that the IMBH-BH binary may be kicked out by a three-body interaction event, as discussed in the previous section. Equation (\ref{eqn:aej}) shows the typical binary semi-major axis below which a three-body interaction will cause the binary to be ejected from the GC. As also discussed in \citet{hol08}, a simple estimate of the relative importance of the dynamical kicks and GW recoil kicks is not easy and depends on the details of the interaction, hence on the ambient cluster. However, dynamical kicks may be important (under some circumstances) only for low-mass IMBHs. For instance, \citet{gul06} found that a $100\ \mathrm{M}_{\odot}$ IMBH is ejected $\approx 50\%$ of the time if its companion and the third BH are $10\ \mathrm{M}_{\odot}$. In our calculations, we neglect the dynamical kicks for simplicity. However, we note that they would affect low-mass clusters that host low-mass IMBH, which are very likely to be ejected soon after the cluster formation because of GW recoil kicks.

\section{Results}
\label{sect:res}

We evolve the primordial GC population by means of the equations described in Section \ref{sect:gcev}, according to the prescriptions in \citet{gne14}. While the cluster evolves in the Galactic field, we generate IMBH-SBH merger events once during every $t_{\mathrm{coll}}$ time interval. Following each merger, we calculate the recoil velocity using the equations shown in Section \ref{sect:imbh}, and update the total spin and mass of the merger remnant to account for radiation of angular momentum and energy \citep{lou10,lou12}. We compute the escape velocity from the center of a GC
\begin{equation}
v_{\mathrm{e}}(t)=\sqrt{\frac{G M_{\mathrm{GC}}(t)}{r_{\mathrm{h}}(t)}}\ ,
\end{equation}
where $r_{\mathrm{h}}(t)$ is the half-mass radius at time $t$, computed using Eq. (\ref{eqn:rhalfm}). As also discussed in Sect. \ref{sect:gcev}, \citet{web14} have shown that eccentric orbits may affect the overall properties of GCs, such as their half-mass radius. This would in turn affect the expected cluster escape speed. However, since we do not know the initial eccentricity distribution of GCs and how the cluster eccentricity affects the half-mass radius of GCs as function of time and cluster mass, we simply adopt the eccentricity parameter $f_e$ of \citet{gne14} to crudely account for eccentricity effects, and use the half-mass radius as computed from Eq. \eqref{eqn:rhalfm}. If $|\textbf{v}_{\mathrm{kick}}|>v_{\mathrm{e}}(t)$, we assume that the IMBH escaped from the star cluster. If the IMBH does not escape, we continue to evolve the cluster mass and position self-consistently according to the equations in Section \ref{sect:gcev}, and generate a new IMBH-SBH merger after $\delta t=t_{\mathrm{coll}}$. If $|\textbf{v}_{\mathrm{kick}}|<v_{\mathrm{e}}(t)$ after $N_{\mathrm{coll}}$ merger events and the IMBH is not ejected from the star cluster, two outcomes are viable. First, the cluster may be dissolved by the Galactic tidal field. We define the IMBH in this channel to have a dissolved GC host. Second, the GC survives and the IMBH remains in the cluster core. To summarize, there are three possible outcomes for each IMBH in a GC
\begin{itemize}
\item the IMBH is ejected from the star cluster as a consequence of the recoil velocity kick;
\item the IMBH is not ejected due to collisions, but the cluster is dissolved by evaporation and/or tidal disruption in the galaxy;
\item the IMBH is retained in the cluster and the cluster remains intact until present.
\end{itemize}

As discussed in the previous Section, several parameters play a role on the fate of the central IMBH. To check the effect of each of them, we run $5$ different models. Table \ref{tab:models} summarizes all the models considered in this work. Every result is the average over $\mathcal{N}=1000$ realizations.

\begin{table}
\caption{Models: name, initial fraction of cluster mass in IMBH ($f$), spin ($\chi$), slope of the SBH mass distribution ($\zeta$), eccentricity ($e$), number of collisions ($N_{\mathrm{coll}}$), time between collisions ($t_{\mathrm{coll}}$), number of realizations ($\mathcal{N}$).}
\centering
\begin{tabular}{lcccccc}
\hline
Name 		& $f$ (\%)	& $\chi$	& $\zeta$	& $e$	& $t_{\mathrm{coll}}$ (Myr) & $\mathcal{N}$\\
\hline\hline
Model 1		& $0.5$-$4$			& $0$		& $1$	 	& $0$ 			& $50$			&1000\\
Model 2		& $1$ 				& $0$ 		& $1$-$4$ 	& $0$ 			& $50$			&1000\\
Model 3		& $1$				& $0$-$0.7$	& $1$	 	& $0$ 			& $50$			&1000\\
Model 4		& $1$				& $0$		& $1$	 	& $0$-$0.2$ 	& $50$			&1000\\
Model 5		& $1$				& $0$		& $1$	 	& $0$			& $50$-$200$	&1000\\
\hline
\end{tabular}
\label{tab:models}
\end{table}

\subsection{IMBH mass distribution in different channels}

\begin{figure} 
\centering
\includegraphics[scale=0.55]{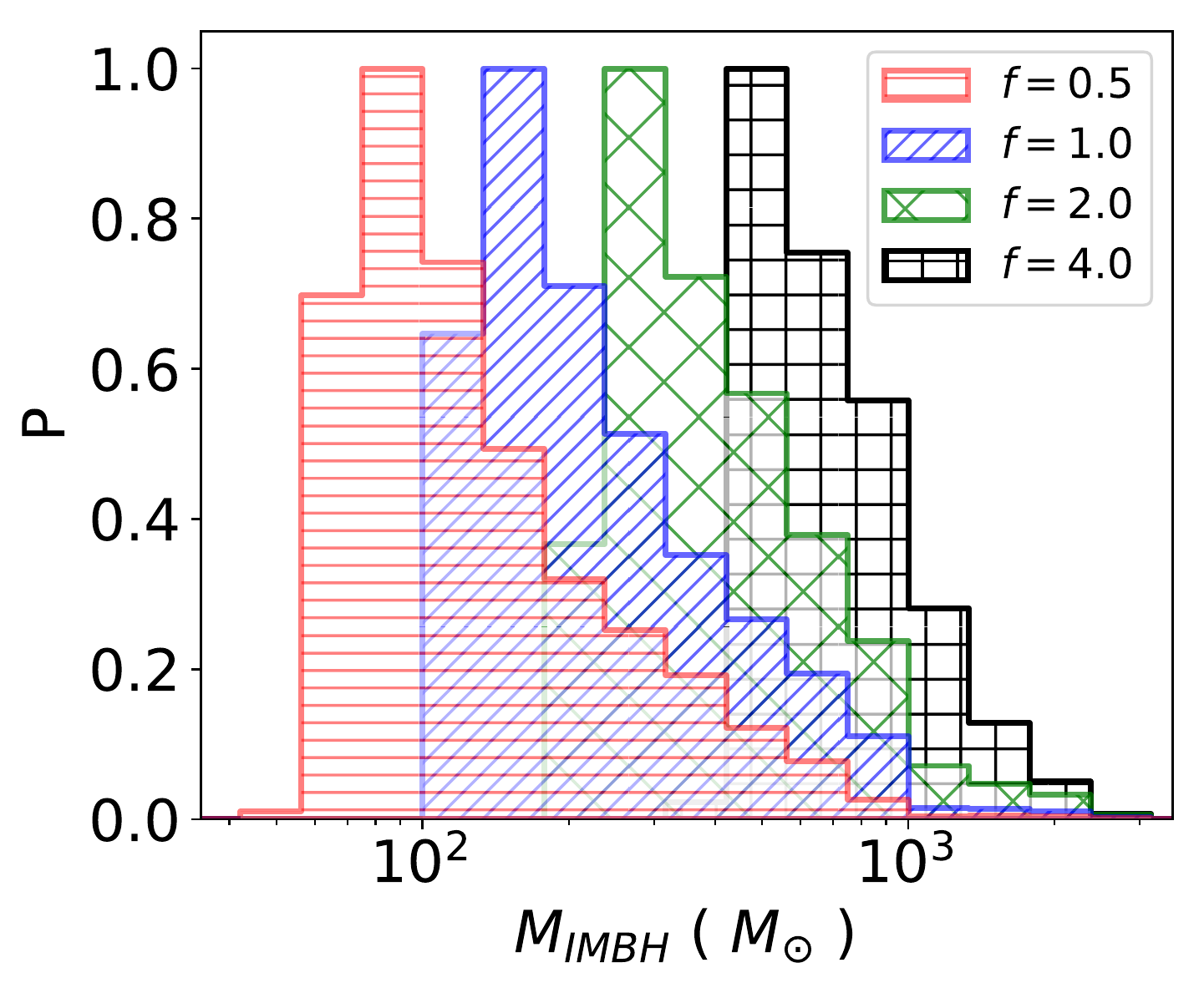}
\includegraphics[scale=0.55]{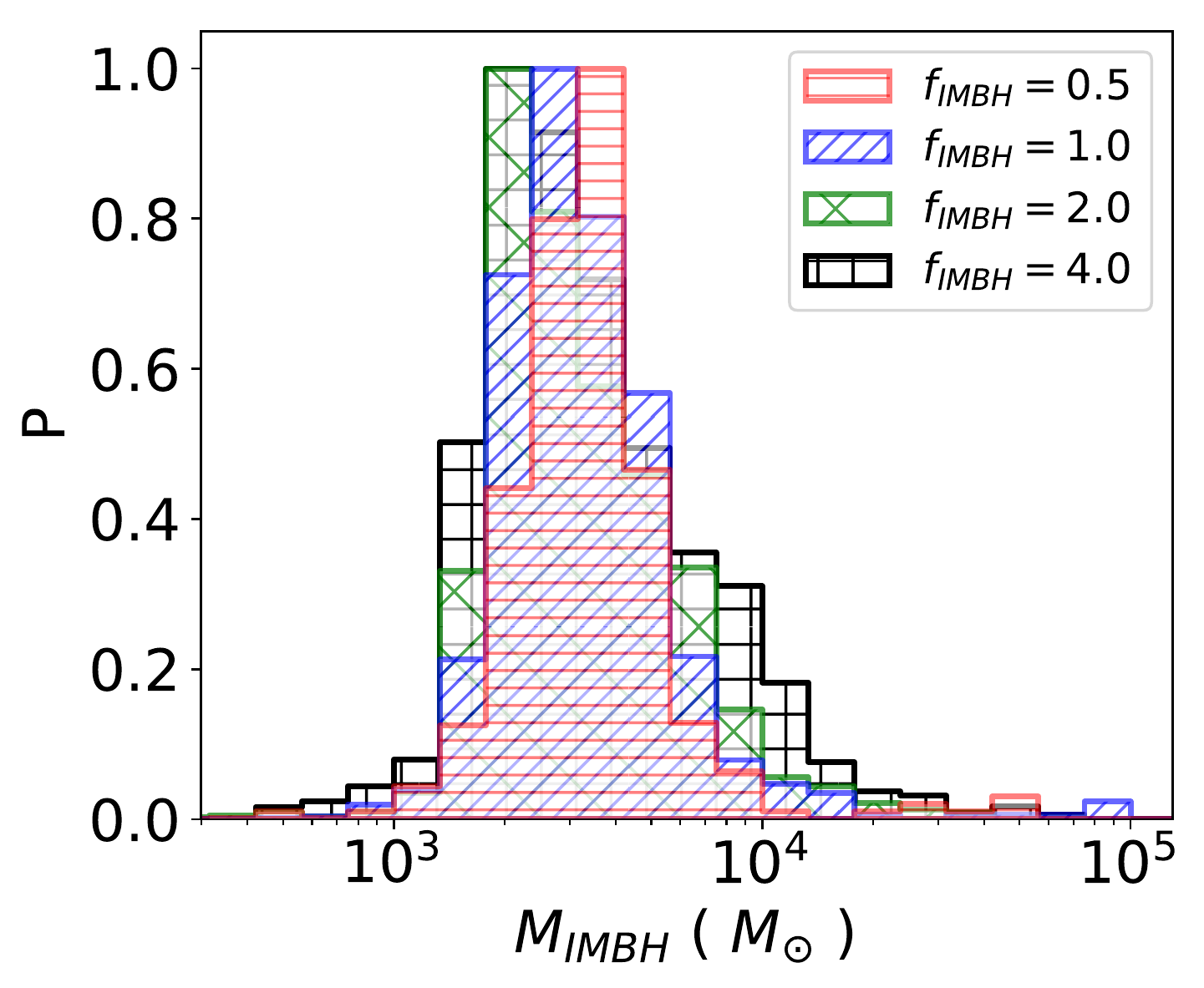}
\includegraphics[scale=0.55]{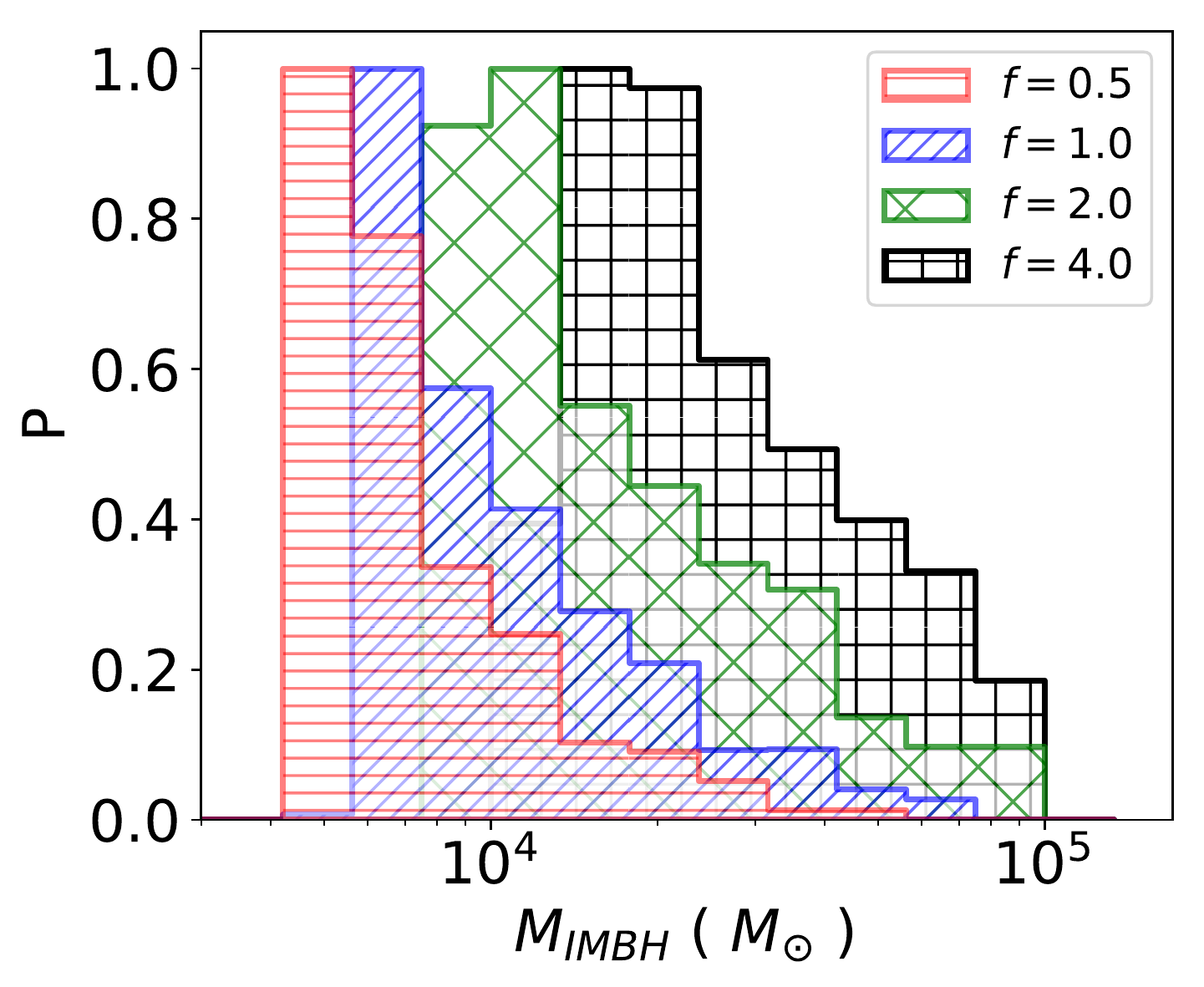}
\caption{IMBH mass distribution for various initial values $f=M_{\mathrm{IMBH}}/M_{\rm GC}$ (shown in the legend in $\%$). The three panels show three different outcomes: IMBH ejected {(\it top panel)}, (ii) IMBHs with dissolved GC host {(\it center panel)}, and (iii) IMBH retained {(\it bottom panel)}. The relative fraction of events for the three different outcomes is shown in Table~\ref{tab:res_fimbh}.}
\label{fig:fimbh_mass}
\end{figure}

\begin{table}
\caption{Branching ratios of different channels and average number of mergers per GC as a function of the initial $f=M_{\mathrm{IMBH}}/M_{\rm GC}$}.
\centering
\begin{tabular}{ccccc}
\hline
$f$ (\%)	&	Ejected (\%)	&	Dissolved (\%)	& Retained (\%)	& $\langle N \rangle$	\\
\hline\hline
$0.5$		& $93.0$		& $4.2$	 	& $2.8$ 			& $13.2$	\\
$1$ 		& $88.3$		& $8.9$ 	& $2.8$ 			& $17.5$	\\
$2$			& $80.2$		& $17.0$ 	& $2.8$ 			& $22.8$	\\
$4$			& $67.0$		& $30.2$ 	& $2.8$			 	& $28.0$	\\
\hline
\end{tabular}
\label{tab:res_fimbh}
\end{table}

\begin{figure*} 
\centering
\begin{minipage}{18cm}
\includegraphics[scale=0.55]{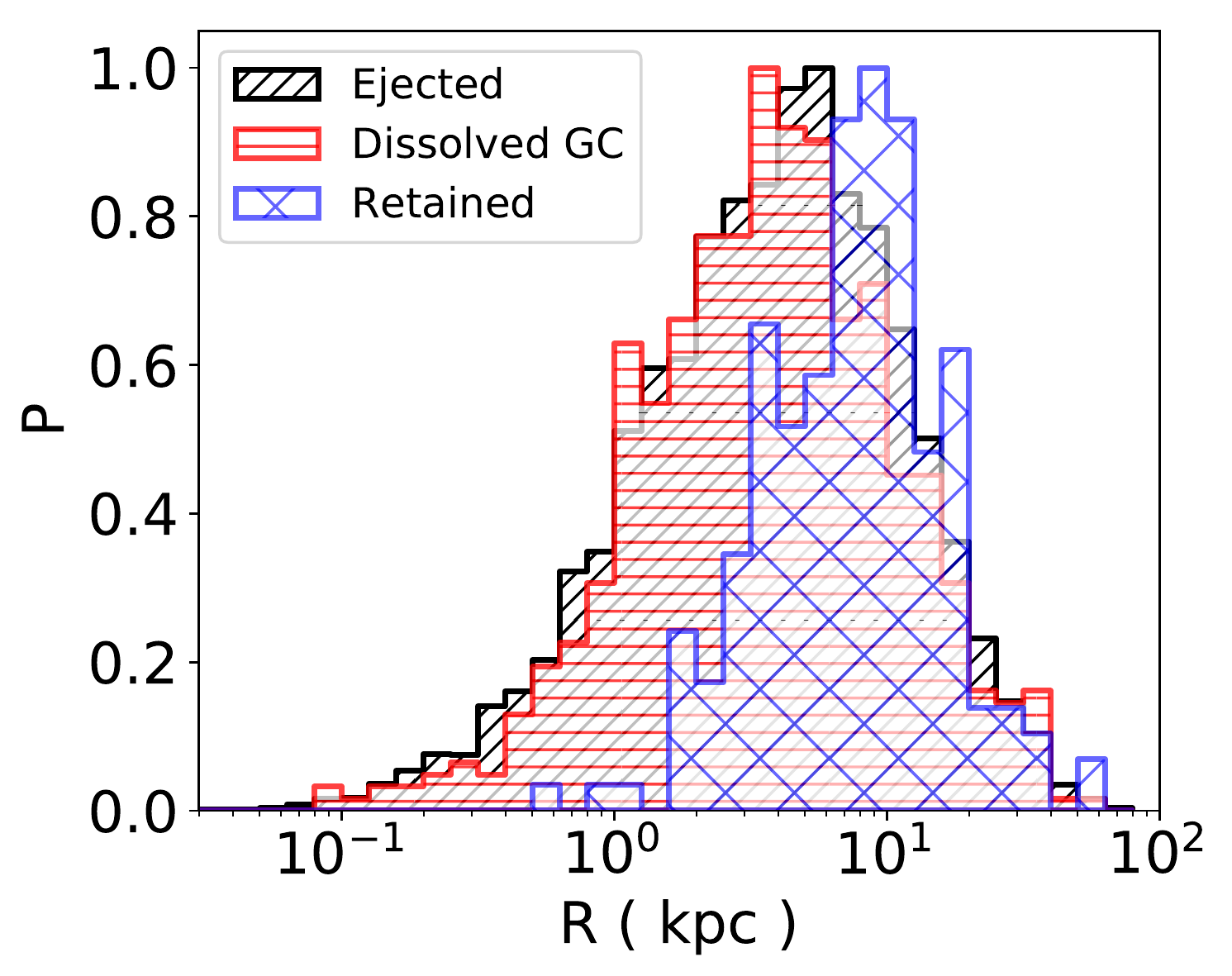}
\hspace{1cm}
\includegraphics[scale=0.55]{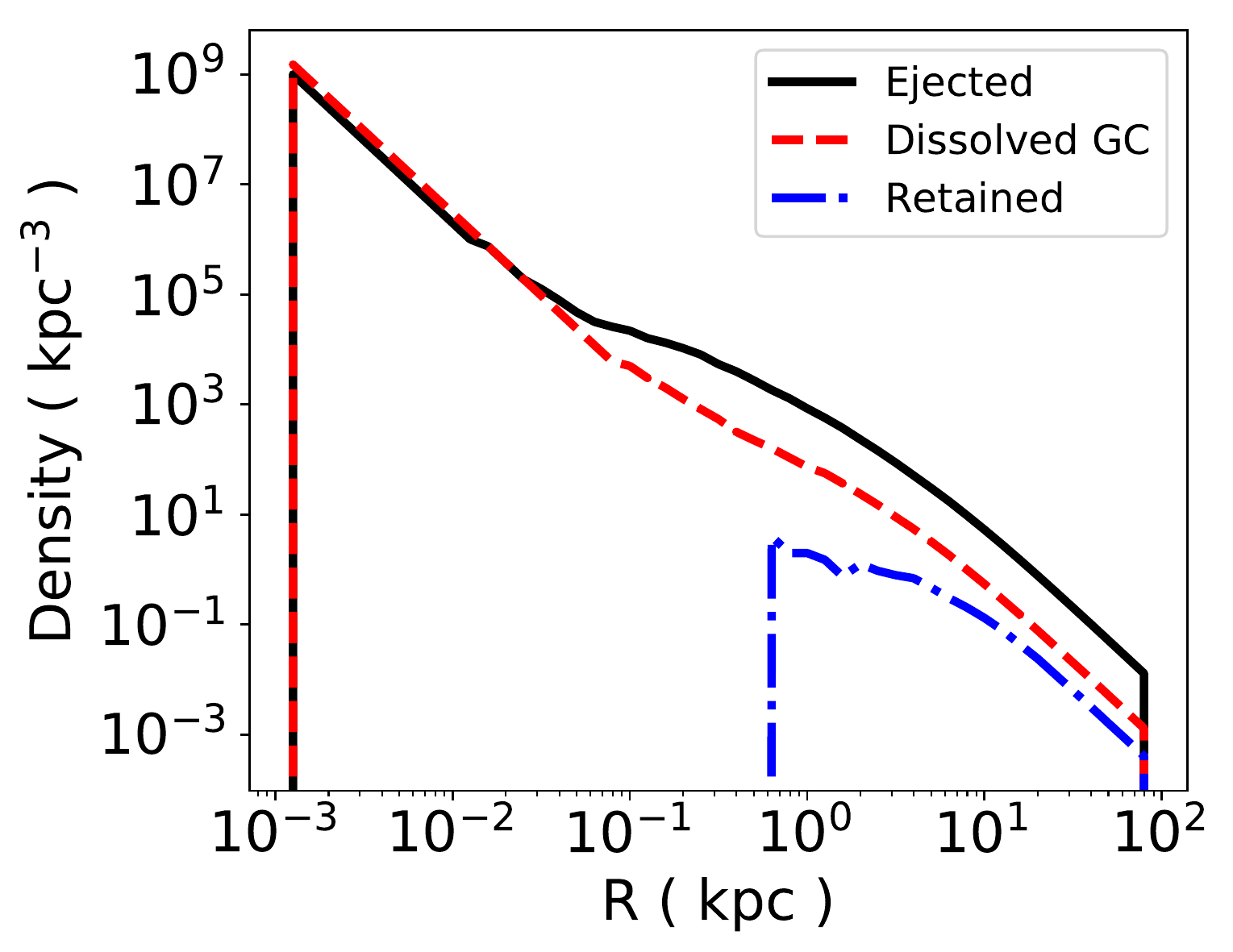}
\caption{Left panel: the final spatial distribution of the IMBH in the three different channels for Model 1 where the fraction of the GC mass in IMBH is $f=1$\%. Right panel: the final spatial IMBH number density in the three different channels for Model 1 when the fraction of the GC mass in IMBH is $f=1$\%.}
\label{fig:fimbh_spatial}
\end{minipage}
\end{figure*}

\begin{figure} 
\includegraphics[scale=0.55]{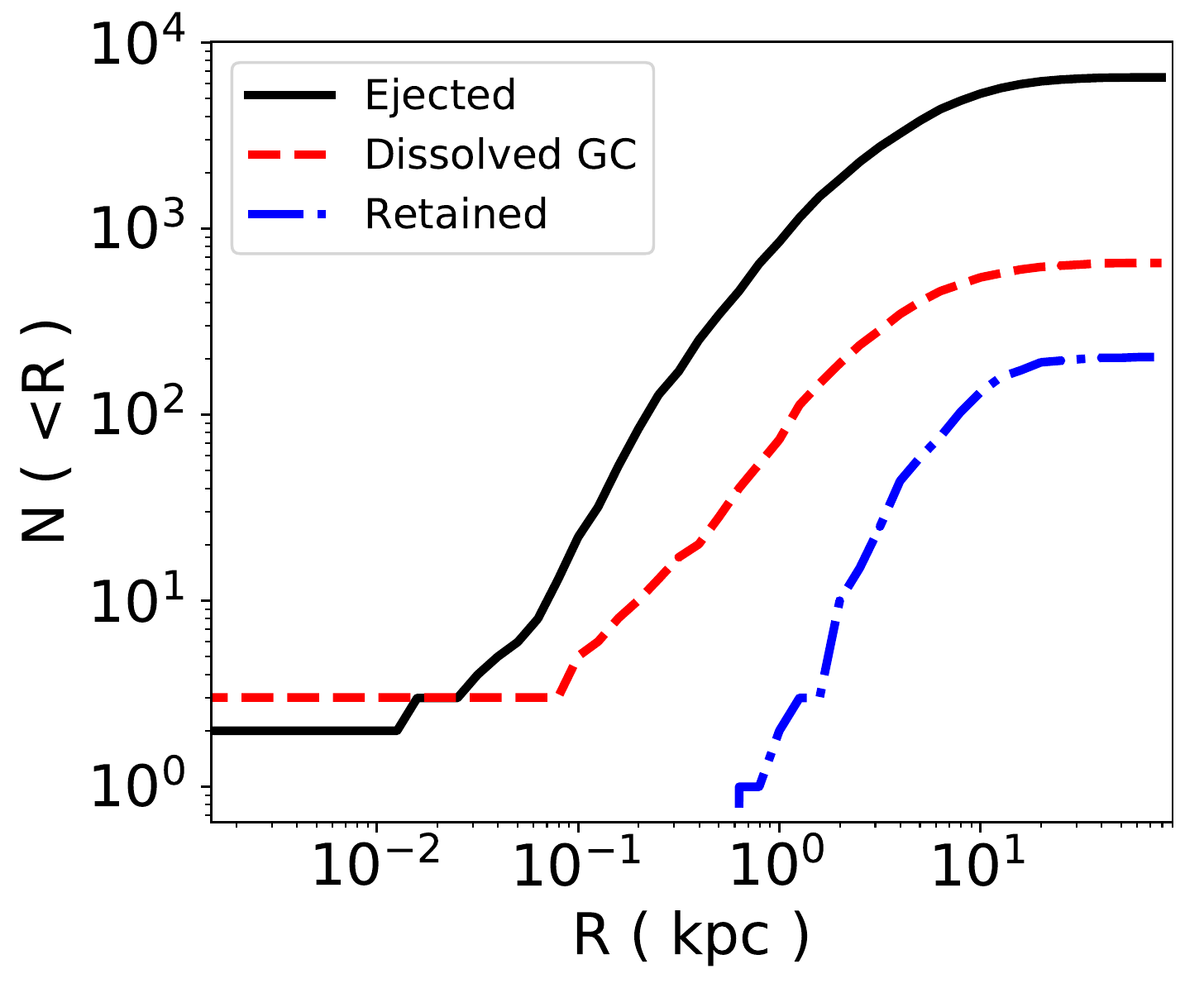}
\caption{Total number of IMBH within a distance $R$ from the galactic center in the three different channels for Model 1 when the fraction of the GC mass in IMBH is $f=1$\%.}
\label{fig:fimbh_number}
\end{figure}

In Model 1 (see Table~\ref{tab:models}), we vary 
the initial mass of the IMBH relative to the initial mass of the cluster, $f=M_{\mathrm{IMBH}}/M_{\rm GC}$ between $0.5\%$ and $4\%$. Since $M_{\mathrm{GC},\min}=10^4\ \mathrm{M}_{\odot}$ and $M_{\mathrm{GC},\max}=10^7\ \mathrm{M}_{\odot}$, the minimum and maximum mass of the IMBHs are $50\ \mathrm{M}_{\odot}$ and $5\times 10^4\ \mathrm{M}_{\odot}$ respectively in the case $f=0.5\%$, while it is $400\ \mathrm{M}_{\odot}$ and $4\times 10^5\ \mathrm{M}_{\odot}$ when $f=4\%$, respectively.  

In this model, both the IMBH and the SBHs have no initial spin. SBH mass is sampled from a distribution $dN/dm = m^{-1}$, i.e.  $\zeta=1$ in Equation~(\ref{eqn:mass}) and all the merger events are assumed to have $e=0$. In total, we generate at most $N_{\mathrm{coll}}=230$ merger events separated by time intervals of $t_{\mathrm{coll}}=50$ Myr. As noted, $N_{\mathrm{coll}}$ represents the maximum number of collisions that an IMBH can undergo. If the IMBH is ejected or the host GC dissolves due to tidal disruption, the number of mergers will be smaller than $N_{\mathrm{coll}}$.

Table \ref{tab:res_fimbh} reports the relative fraction of systems with different outcomes: IMBHs ejected from the GC, IMBHs whose host GC dissolves, and IMBHs which remain within their host GC which survives until present. The table also displays the average number of mergers per GC. We find that most of the IMBHs are ejected from the star cluster. IMBHs are retained only in the massive GCs and their relative fraction does not depend on $f$. Even for $f=0.5\%$, the IMBHs in these clusters experience very small recoil velocities due to very small mass ratios and the GCs have very large escape velocities. By increasing $f$, hence $M_{\rm IMBH}$, the probability of IMBHs to be ejected from the cluster by present decreases, while the fraction of IMBHs with dissolved hosts becomes larger. 

Figure \ref{fig:fimbh_mass} shows the resulting distributions for the mass of the IMBH in the three different channels as a function of $f$. 
The top panel illustrates the masses of IMBHs that are ejected from their parent clusters as a consequence of the GW merger kicks. The larger $f$, the larger the peak mass of the distribution. Even for $f=4\%$, the mass distribution of ejected IMBHs is peaked at $\lesssim 1000\ \mathrm{M}_{\odot}$. The majority of the IMBHs in this channel comes from low-mass clusters. Since the initial GC mass distribution has a negative power-law (see Eq. \ref{eqn:gcmassini}), most of the clusters have mass $\lesssim 10^5\ \mathrm{M}_{\odot}$ and, as a consequence, most of the IMBHs have mass $\lesssim 10^3\ \mathrm{M}_{\odot}$. For this population, the recoil velocity is large compared to IMBHs in more massive clusters because of their smaller mass ratios (Eq. \ref{eqn:vkick}), and their host GCs have the lowest escape velocities because of their light masses. As a consequence, the IMBHs in this regime are ejected from their parent clusters soon after cluster formation. Finally, we note that such ejected IMBHs may even escape their host galaxy as a consequence of the GW velocity kick. However in case the BH spin is zero, the kick velocity is typically not large enough to overcome the Galactic potential well, as it is at maximum of the order of $\approx 200$ km s$^{-1}$ in this case. When the effect of the spin is taken into account, the kick velocity may be as large as few thousands km s$^{-1}$ and IMBHs may escape their host galaxy (see Section~\ref{sect:spins}). 

The central panel of Fig. \ref{fig:fimbh_mass} illustrates the mass of the IMBHs with dissolved GC host. The distributions are peaked at $\approx 1,000$--$3,000\ \mathrm{M}_{\odot}$ and are roughly independent of the GC mass fraction in IMBH. The IMBHs that belong to this class were born in GCs of intermediate masses. In these GCs, the recoil velocity is smaller than for the ejected IMBHs as a consequence of the smaller $q$ and the escape velocity from the cluster is larger as a consequence of an initial larger mass. Hence, the IMBHs hosted in such clusters tend to be retained without being ejected after a few collisions. We also note a tail of the $M_{\mathrm{IMBH}}$ distribution up to $\approx 10^5\ \mathrm{M}_{\odot}$. These IMBH were members of the most massive GCs, the ones with mass near to $M_{\mathrm{GC},\max}=10^7\mathrm{M}_{\odot}$. Due to their short dynamical friction timescales (i.e. a few Gyr for $M_{\mathrm{GC}}\gtrsim 10^6\ \mathrm{M}_{\odot}$), such clusters are transported to the galactic center and are tidally disrupted by the Galactic gravitational field (Eq. \ref{eqn:dens}). In our model with $f=1\%$, we find that $\approx 0.25$\% of the primordial IMBH population with mass $\gtrsim 10^4\ \mathrm{M}_{\odot}$ was hosted by GCs of mass $\gtrsim 10^6\ \mathrm{M}_{\odot}$ dissolved by the Galactic field.

\begin{figure} 
\includegraphics[scale=0.55]{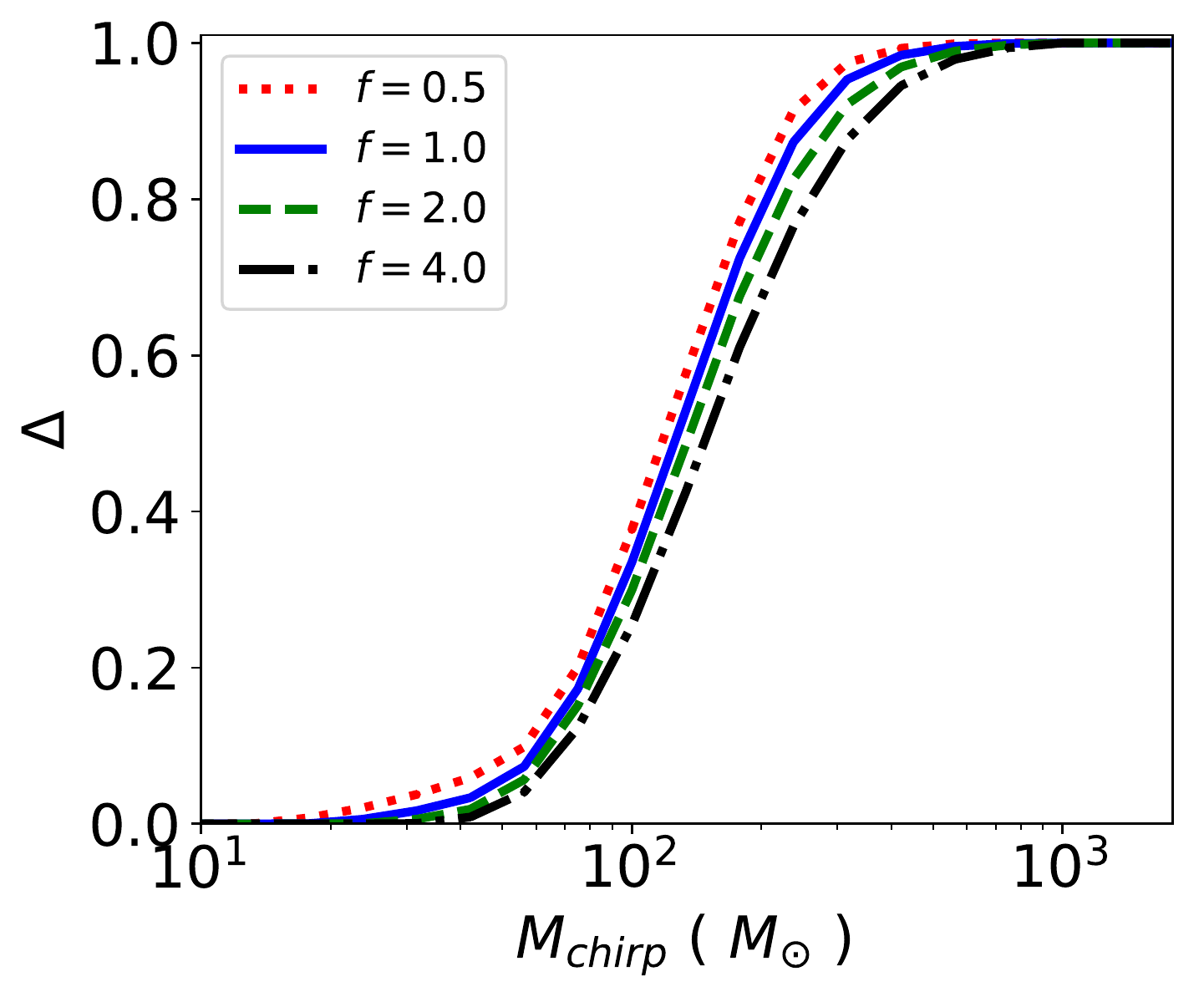}
\includegraphics[scale=0.55]{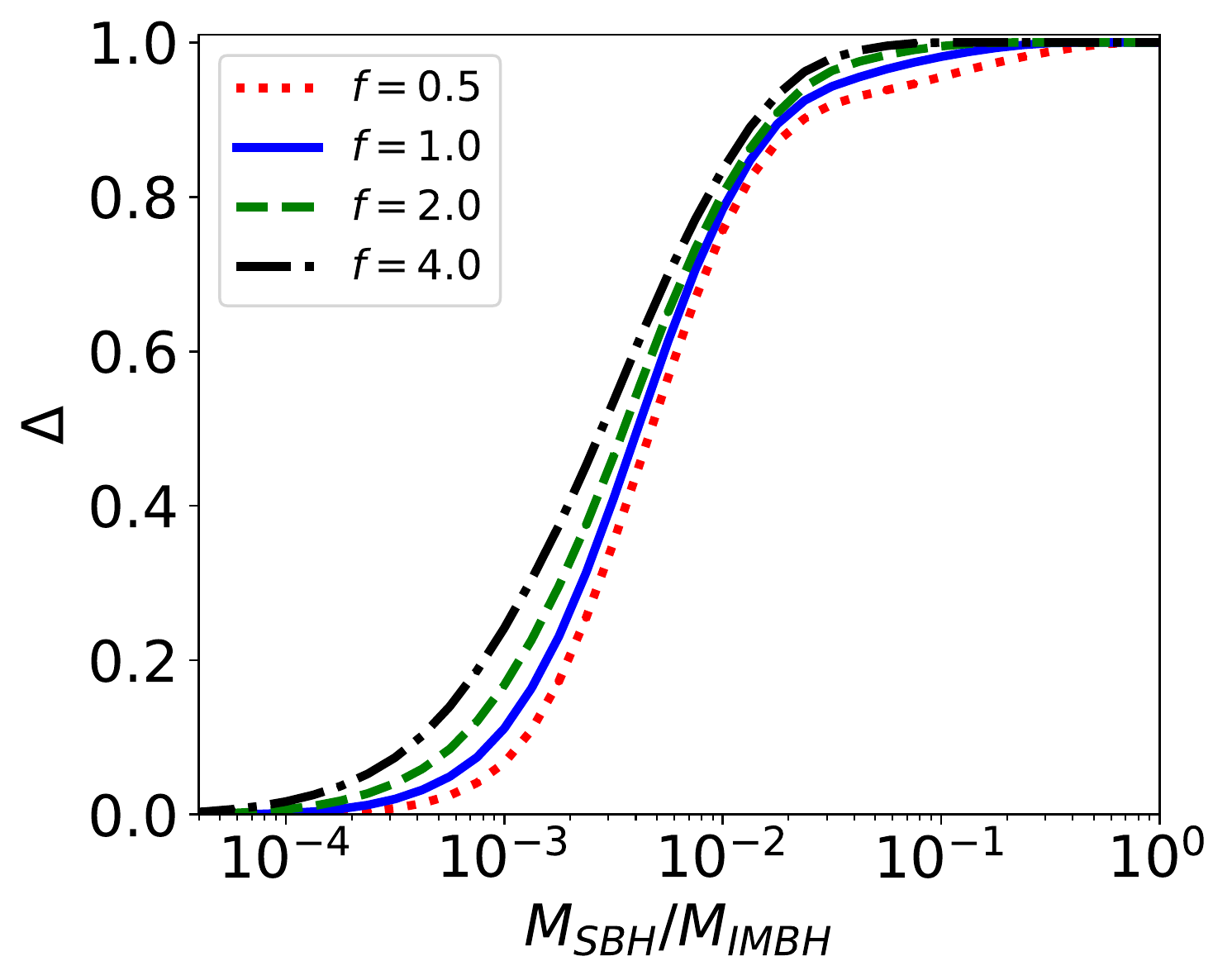}
\caption{Chirp mass (top) and mass ratio (bottom) cumulative functions of the IMBH-SBH merger events as a function of the fraction $f$ of the cluster mass in IMBH.}
\label{fig:fimbh_chirp}
\end{figure}

\begin{figure} 
\includegraphics[scale=0.55]{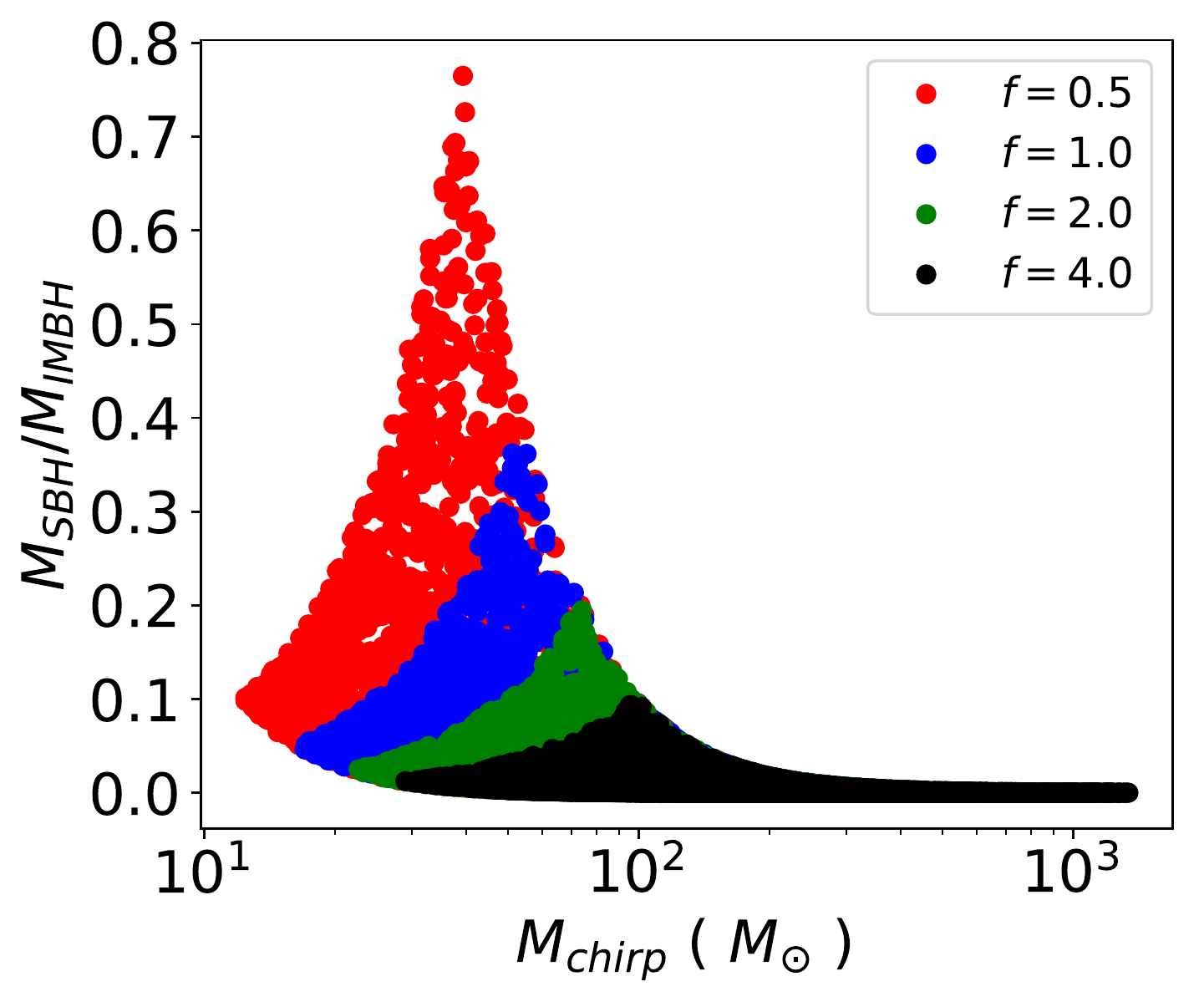}
\caption{Mass ratio as a function of the IMBH-SBH merger events for different fractions $f$ of the cluster mass in IMBH.}
\label{fig:fimbh_ratiomtot}
\end{figure}

The bottom panel of Fig. \ref{fig:fimbh_mass} shows the mass of IMBH retained in GCs. The peak of the distribution is at $4,000$--$20,000\ \mathrm{M}_{\odot}$ depending on $f$ with larger peak $M_{\mathrm{IMBH}}$ values corresponding to larger $f$. IMBHs in this channel are in the more massive GCs. Such very massive IMBHs can survive for all the $N_{\mathrm{coll}}$ merger events until present mainly for two reasons. First, unlike low-mass IMBHs, these heavier IMBHs are imparted very small kick velocities because of the very unequal mass ratios. Second, the escape velocity from the parent cluster core is relatively large because of the large host GC mass. However, as discussed previously, IMBHs from the extremely massive GCs are not retained in intact GCs as those sink efficiently to the centers of the host galaxies and are disrupted by the strong gravitational field therein.

We note that based on the bottom panel, the IMBH candidate of $\approx 40,000\ \mathrm{M}_{\odot}$ detected in $\omega$ Cen \citep{bau17} is most probable for initial values $M_{\mathrm{IMBH}}/M_{\mathrm{GC}}\sim 4\%$ and $M_{\mathrm{IMBH}}/M_{\mathrm{GC}}\lesssim 1\%$ are highly disfavored. However the $\approx 2,200\ \mathrm{M}_{\odot}$ IMBH in $47$ Tuc \citep{kiz17}, is only possible for initial values $M_{\mathrm{IMBH}}/M_{\mathrm{GC}}\lesssim 0.5\%$. 

\subsection{Radial distribution of IMBHs in the host galaxy}

For the three outcomes (IMBH ejection, GC dissolution, IMBH retention), we have evaluated the final spatial distribution of IMBHs. For all the IMBHs in the first two outcomes, we estimate the final spatial distribution by evolving their orbit by means of dynamical friction (see Eq. \ref{eqn:dynf}). As discussed, all the ejected IMBH are retained by the host galaxy if they do not have spin. In such a case, the total number of IMBHs is of the order of the primordial population of GCs ($\approx 7000$ for the Milky Way). Figure \ref{fig:fimbh_spatial} illustrates the spatial distribution and the number density of the IMBHs in the three different channels as a function of the radial distance from the galactic center. The IMBH surviving in clusters maps the spatial distribution of the GC population (see \citealt{gne14,fao17}). They are concentrated at $\approx 10$ kpc, where the Galactic tidal field was weak enough to not overcome the self-gravity of their parent clusters. However, IMBHs in the first two channels are peaked at smaller distances ($\approx 5$-$6$ kpc) and their distributions have similar shapes. In the case of IMBHs that are not ejected until present, but are in GCs disrupted by the galaxy, the smaller distances are due to the fact that clusters are more efficiently destroyed where the galaxy's gravitational field is stronger. On the other hand, IMBHs that are ejected as a consequence of the recoil kick velocity are concentrated at small distances since the primordial GC spatial profile maps the Galactic star distribution and most of the clusters have masses $\lesssim 10^5\ \mathrm{M}_{\odot}$. Figure \ref{fig:fimbh_number} illustrates the number of IMBHs within a distance $R$ from the galactic center. Our models predict $\approx 1000$ IMBH within $1$ kpc.

\subsection{Mass distribution of IMBH-SBH mergers}

\begin{figure} 
\centering
\includegraphics[scale=0.55]{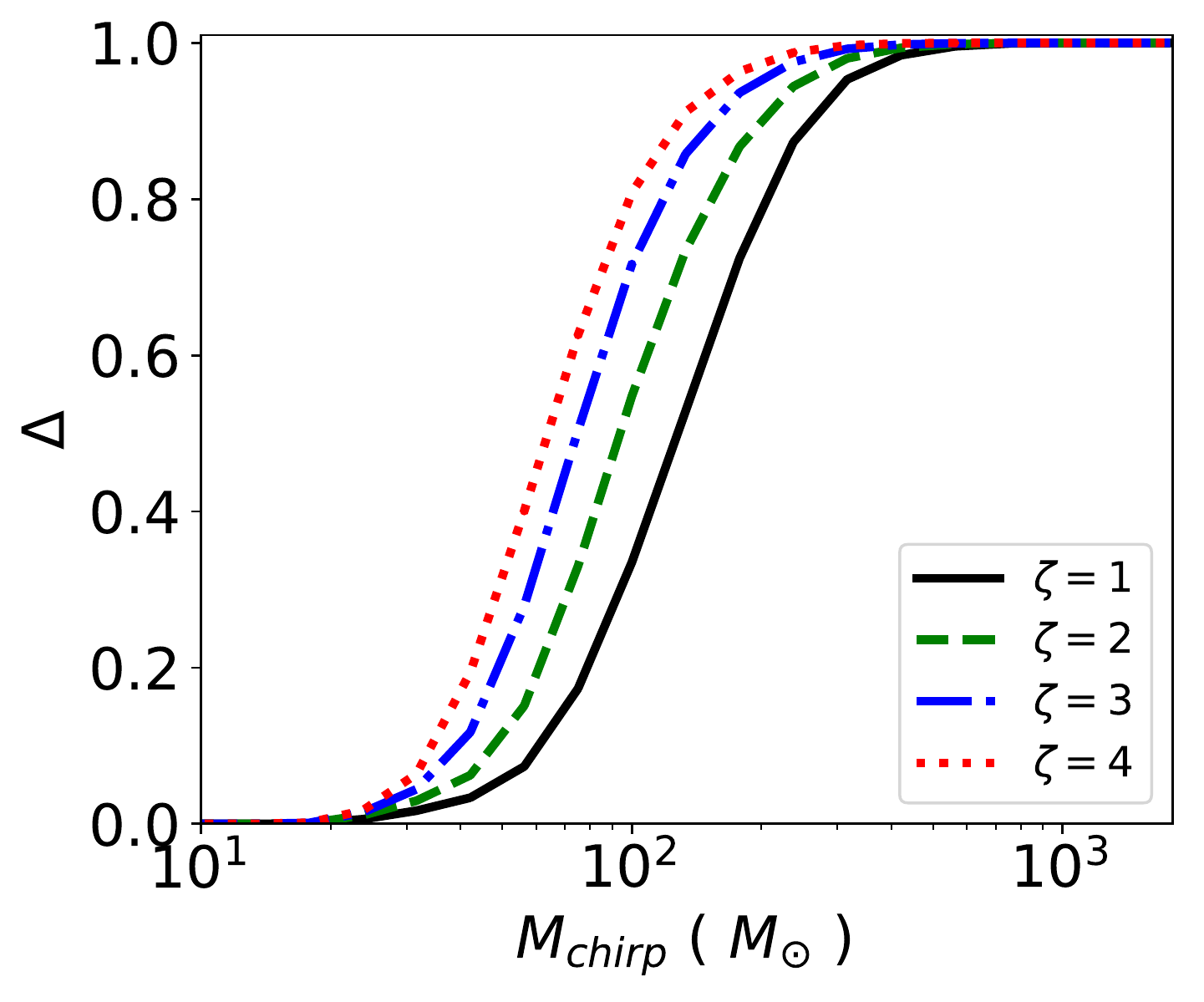}
\includegraphics[scale=0.55]{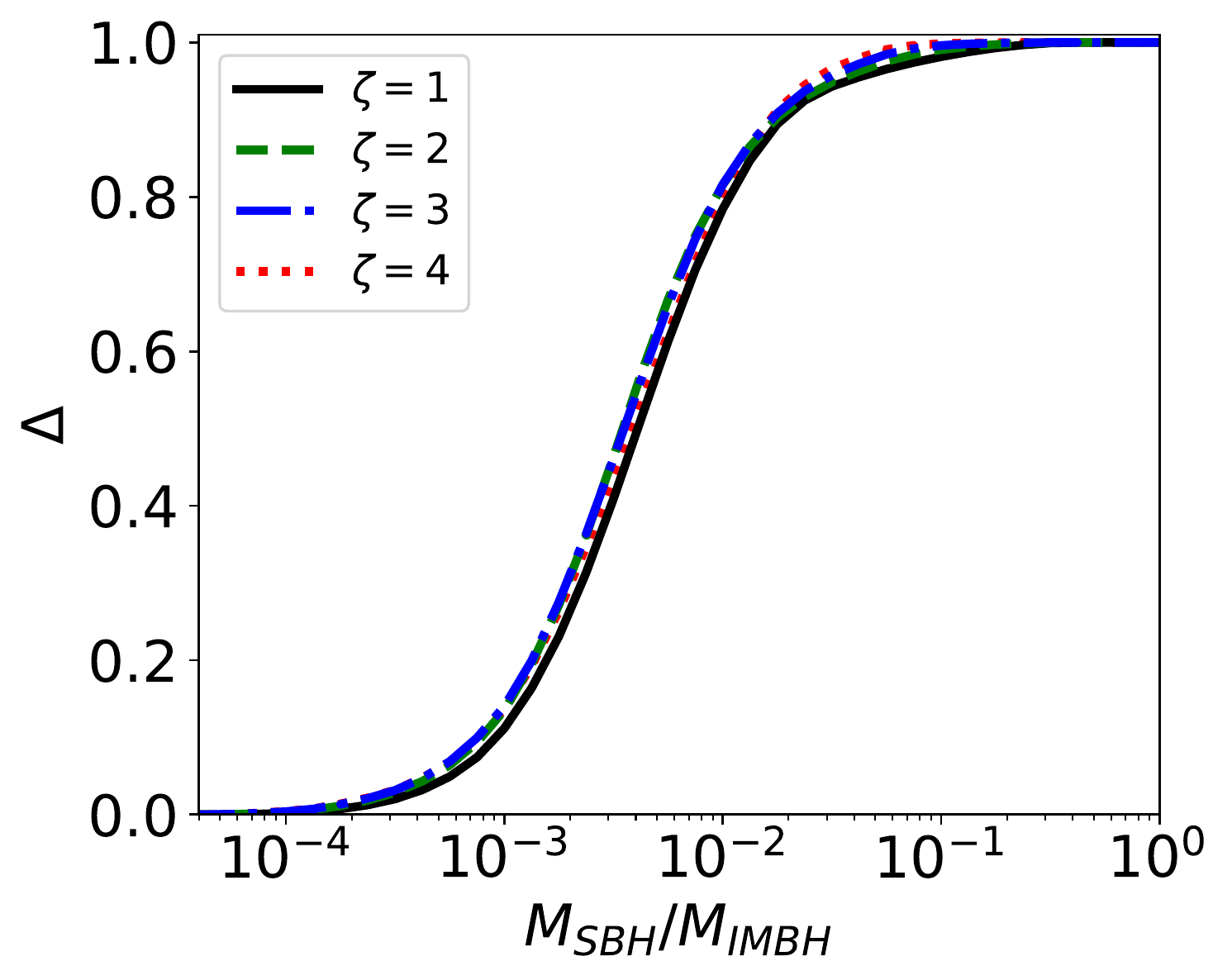}
\caption{Effect of the slope $\zeta$ of the SBH mass function: chirp mass cumulative function (top) and mass ratio cumulative function (bottom).}
\label{fig:beta_chirp}
\end{figure}

\begin{figure} 
\centering
\includegraphics[scale=0.55]{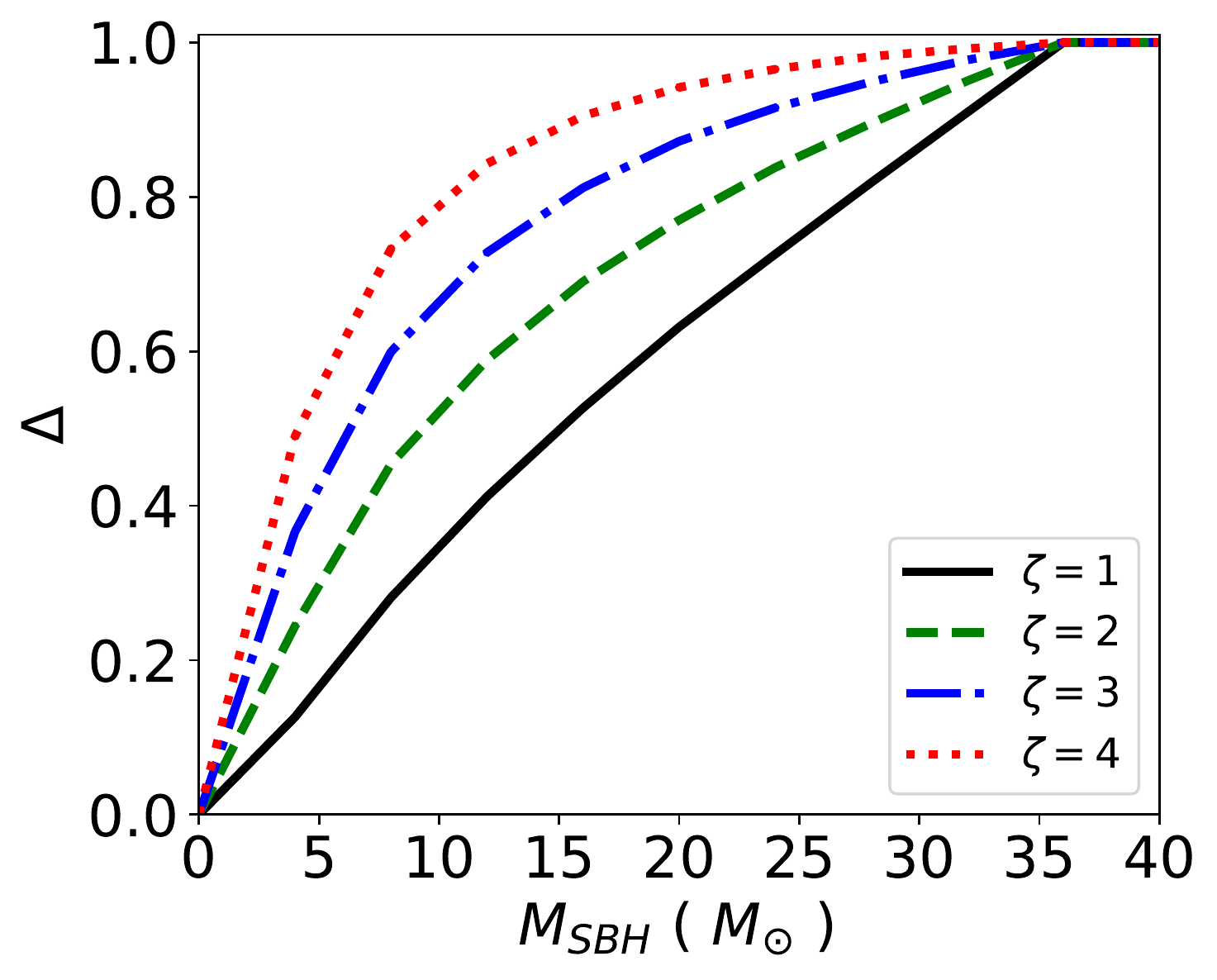}
\caption{Among clusters in which IMBHs are ejected, the cumulative distribution of the mass of the stellar black hole during the last merger leading to IMBH ejection for various SBH mass funtion slope $-\zeta$. }
\label{fig:beta_lastsbh}
\end{figure}

Two of the most interesting quantities that can be measured during a GW event (see Section \ref{sect:imris}) are the chirp mass ($M_{\mathrm{chirp}}=(M_{\mathrm{IMBH}}\ M_{\mathrm{SBH}})^{3/5}/(M_{\mathrm{IMBH}}+M_{\mathrm{SBH}})^{1/5}$) and the mass ratio ($q=M_{\mathrm{SBH}}/M_{\mathrm{IMBH}}$). Figure \ref{fig:fimbh_chirp} shows the cumulative distributions $\Delta$ of chirp mass and mass ratio of all the IMBH-SBH merger events as a function of the initial IMBH mass relative to the GC mass, $f$ for Model 1. Small $f$'s imply smaller chirp masses and larger mass ratios. In the case $f=0.5\%$, approximately $50\%$ of mergers have chirp mass $\le 120\ \mathrm{M}_{\odot}$, which fraction reduces to $30$ \% for $f=4\%$. When $f=0.5\%$, $50\%$ of mergers have $q\lesssim 5\times 10^{-3}$, while $50\%$ of mergers have $q\lesssim 3\times 10^{-3}$ if $f=4\%$. Figure \ref{fig:fimbh_ratiomtot} shows the distribution of mass ratios as a function of the chirp mass for different $f$'s. For $f=0.5\%$, chirp masses of $50\ \mathrm{M}_{\odot}$ may have mass ratios up to $\approx 0.8$, while for $f=4\%$ the mass ratio is $\lesssim 0.1$ independently of the chirp mass of the IMBH-SBH binary.

In Model 2 (see Table~\ref{tab:models}), we fix the fraction $f=1\%$ of the initial cluster mass in IMBH, while SBH masses are sampled from a power-law distribution with negative exponent $\zeta$ that we vary between $1$ and $4$ (Equation~\ref{eqn:mass}). Figure \ref{fig:beta_chirp} shows the cumulative distribution of the chirp mass and the cumulative distribution $\Delta$ of the mass ratio as a function of $\zeta$. Larger $\zeta$ implies a smaller number of SBHs with large masses, which imply larger chirp masses. The shallower the slope the larger the typical chirp mass of the IMBH-SBH merger event. On the other hand, the mass ratio is nearly independent on the slope of the SBH mass distribution.

The slope of the SBH mass distribution $\zeta$ also affects the mass of the SBH in the last IMBH-SBH event leading to the ejection due to GW recoil kick. Figure \ref{fig:beta_lastsbh} reports the cumulative distribution of the mass of the SBHs responsible for IMBH ejections in the first channel. In the case of $\zeta=1$, approximately $50$\% of SBHs of the last merger have mass $\lesssim 16\ \mathrm{M}_{\odot}$, while $50$\% of SBH of the last merger have mass $\lesssim 4\ \mathrm{M}_{\odot}$ for $\zeta=4$. Shallower SBH mass functions (smaller $\zeta$) imply a larger number of massive SBH, hence a larger mass ratio and a higher recoil velocity. As a consequence, the probability of ejected IMBH when $\zeta=4$ is smaller by $\approx 10$\% than the case $\zeta=1$, while the probability of IMBH in the second channel approximately doubles.

\subsection{Spin effects}\label{sect:spins}

\begin{figure} 
\centering
\includegraphics[scale=0.55]{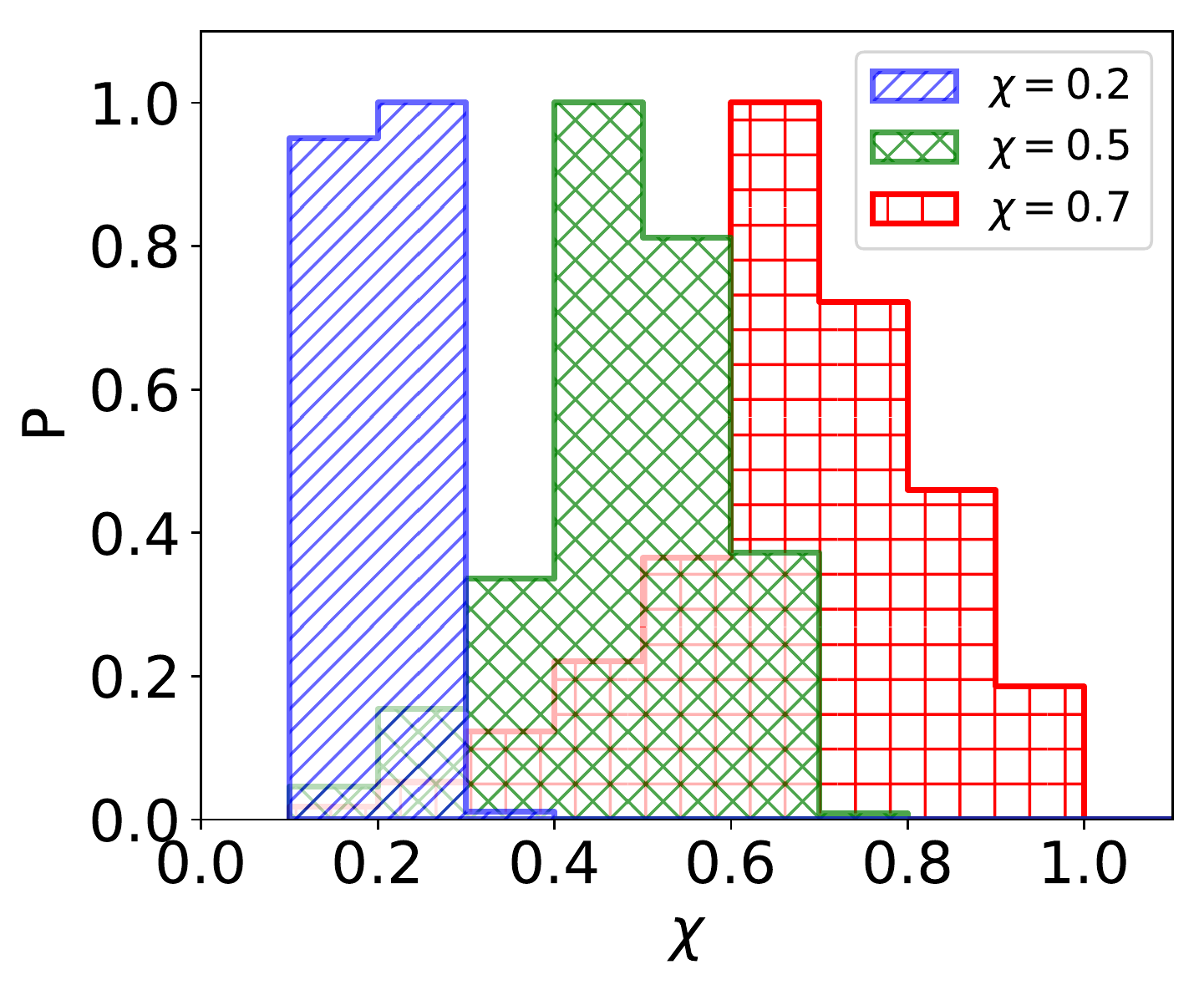}
\caption{Spin distribution of IMBHs for all the merger events as a function of the initial BH spin $\chi$.}
\label{fig:spin_spin}
\end{figure}

\begin{figure} 
\centering
\includegraphics[scale=0.55]{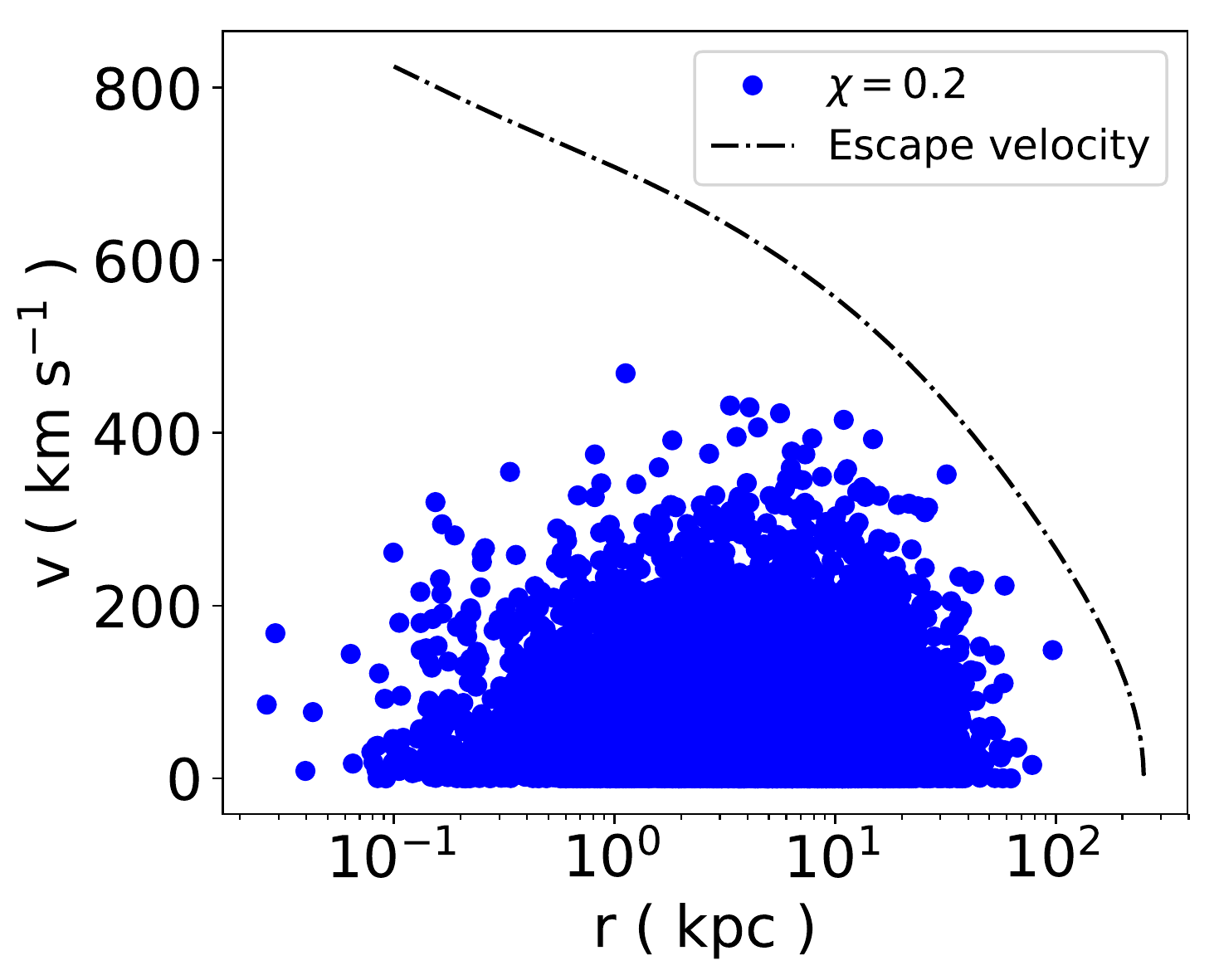}
\includegraphics[scale=0.55]{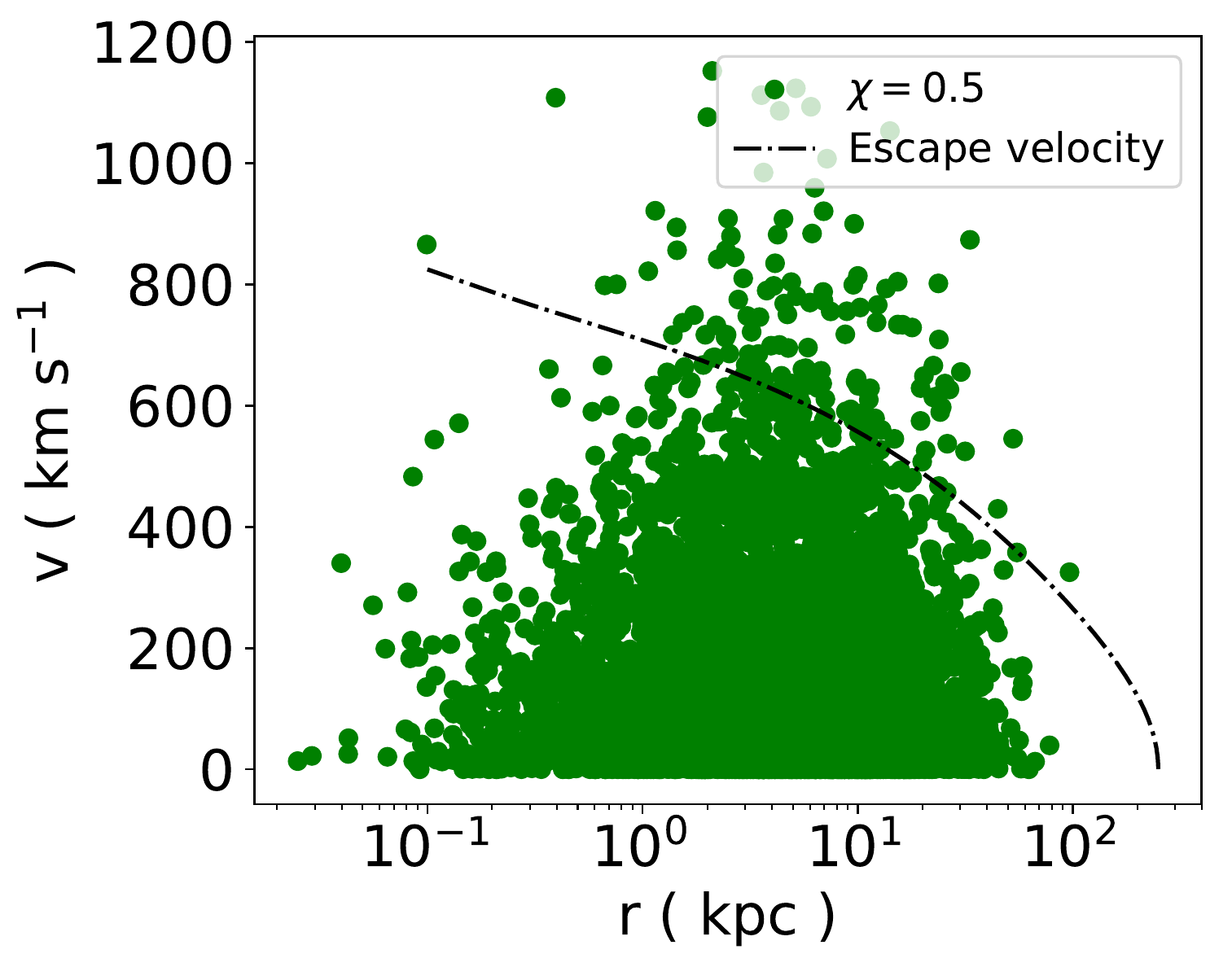}
\includegraphics[scale=0.55]{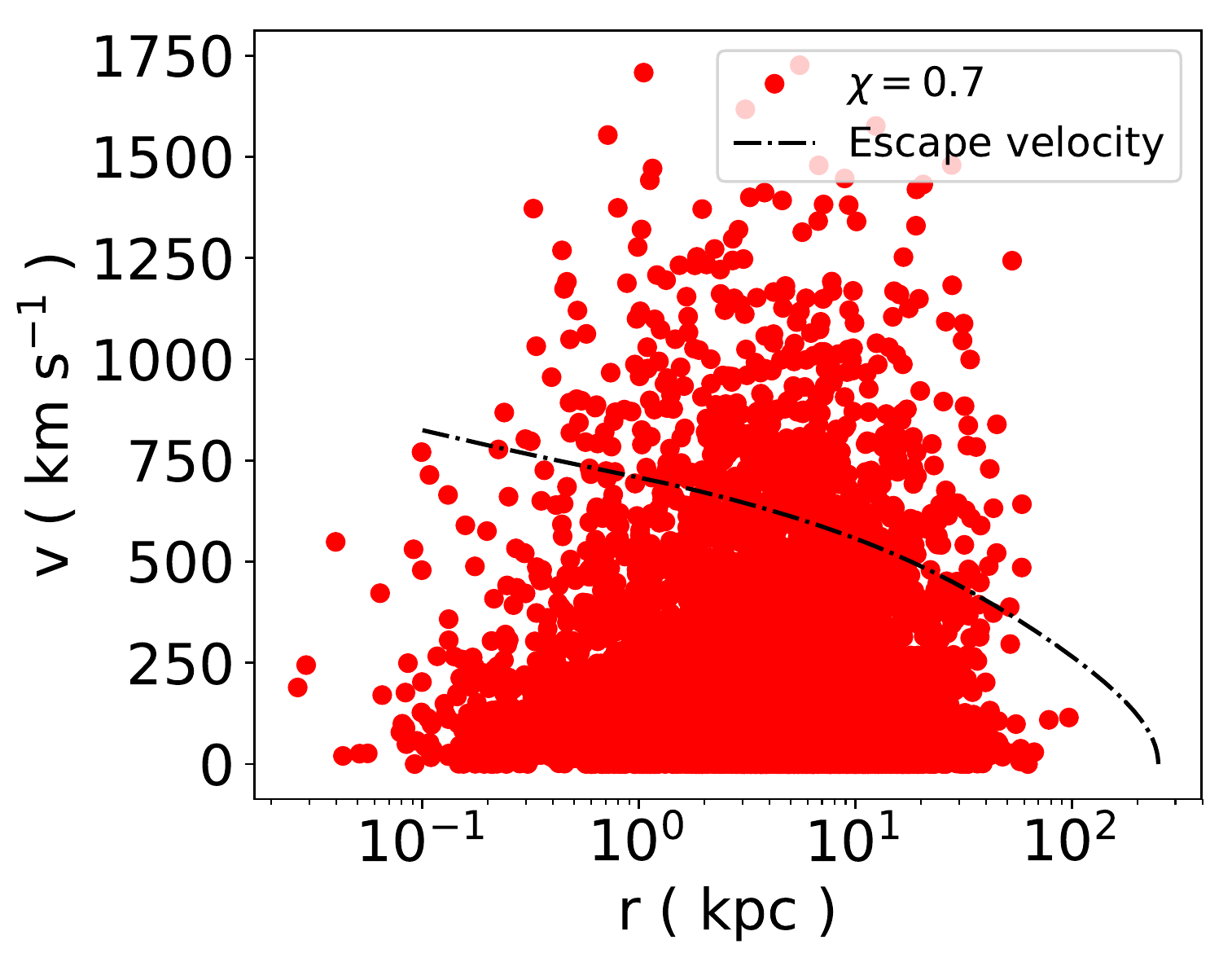}
\caption{Gravitational Wave recoil velocity for all the mergers for different spin of the IMBH and SBH: $\chi=0.2$ (top), $\chi=0.5$ (center), $\chi=0.7$ (bottom).}
\label{fig:escapev_spin}
\end{figure}

In Model 3 (see Table~\ref{tab:models}), we investigate the role of the spin. In this model, both the IMBH and the SBHs have initial reduced spin $\chi$ that we fix to $0.2$, $0.5$, $0.7$, respectively. Different spins are predicted by different physical scenarios. If IMBH mass accumulates by disk accretion, then the spins will be very high, above $\approx 0.9$ \citep{sha05}, if IMBH is created by a large number of mergers it will be peaked at around $0.7$ \citep{fis17}, and if it is formed by a collapse of a very massive star it may be small \citep{ama16,bel17}. The spins of the possible IMBHs hosted in the center of a handful of GCs estimated from the observed radio jet luminosity in the Galaxy are in the range $0-0.35$ \citep{bul11}.

Large spins imply larger recoil velocities. However, the exact outcome depends on the geometry and relative orientation of the IMBH-SBH orbital plane and spin directions. In every merger event, the spin directions are drawn from an isotropic distribution. All the other relevant relative orientations between the spins, angular momenta and orbital plane are generated according to the prescriptions in \citet{lou10}. Figure \ref{fig:spin_spin} shows the distribution of spins of the IMBHs for all the merger events of Model 3. In this model, both the IMBH and SBH have an initial spin $\chi$. After the merging event, we correct the spin of the merger product to account for these losses \citep{lou10}. After $t_{\mathrm{coll}}$, we generate another IMBH-SBH merger, where the IMBH has the spin previously computed and the SBH has spin $\chi$. The resulting distributions of IMRI spins are peaked at nearly the initial $\chi$, with a width that increases with larger values of $\chi$. For equal mass mergers the final spin is peaked at $0.7$ independently of the initial spins due to the orbital angular momentum of the merging BHs \citep{hof16}. In general, the final angular momentum parameter $\chi_f$ of a merger product is the sum of three contributions \citep{buo08}
\begin{equation}
\chi_f=\frac{L_{\mathrm{orb}}(\mu,r_{\mathrm{ISCO}},\chi_f)}{M^3}+\frac{M^3_{\mathrm{IMBH}}\chi_{\mathrm{IMBH}}}{M^3}+\frac{M^3_{\mathrm{SBH}}\chi_{\mathrm{SBH}}}{M^3}\ ,
\label{eqn:finspin}
\end{equation}
where $M$ is the binary total mass, $\chi_{\mathrm{IMBH}}$ and $\chi_{\mathrm{SBH}}$ are the reduced spins of the IMBH and SBH, respectively, and $L_{\mathrm{orb}}(\mu,r_{\mathrm{ISCO}},a_f)$ is the orbital angular momentum of a particle of mass $\mu$ at the ISCO of a Kerr black hole of spin parameter $\chi_{f}$. Being usually $M_{\mathrm{IMBH}}\gg M_{\mathrm{SBH}}$, the spin of the merger product is dominated by the contribution of the IMBH (with a contribution of the angular momentum). This explains the behavior of the distribution of spins in Fig \ref{fig:spin_spin}.

As discussed, in Equation~\eqref{eqn:vkick} the kick velocity is at maximum of the order of $\approx 200$ km s$^{-1}$. An IMBH can be ejected from the host GC, but its kick velocity may not be large enough to overcome also the Galactic potential well when the BH spin is zero. When the effect of the spin is taken into account, the kick velocity can be as large as a few thousands km s$^{-1}$ (the larger the spin the larger the recoil velocity) and IMBHs can escape also their host galaxy. Figure \ref{fig:escapev_spin} shows the GW recoil velocity for different spins of the IMBH and SBHs (Model 3). We find that, while for the model with initial spin parameter of $\chi=0.2$ a negligible fraction of IMBHs escapes the host galaxy, $\approx 2$\% and $\approx 7$\% of the IMBHs ejected from their host GC is also unbounded with respect to the host galaxy in the models with $\chi=0.5$ and $\chi=0.7$, respectively, i.e. $\approx 150$ and $\approx 500$ IMBHs out of the $\approx 7000$ of the initial IMBHs are lost by Milky Way-like type galaxies when $\chi=0.7$. The other parameters, as the fraction of GC mass in IMBH and the slope of the SBH mass function, do not affect the overall result.

\begin{figure} 
\centering
\includegraphics[scale=0.55]{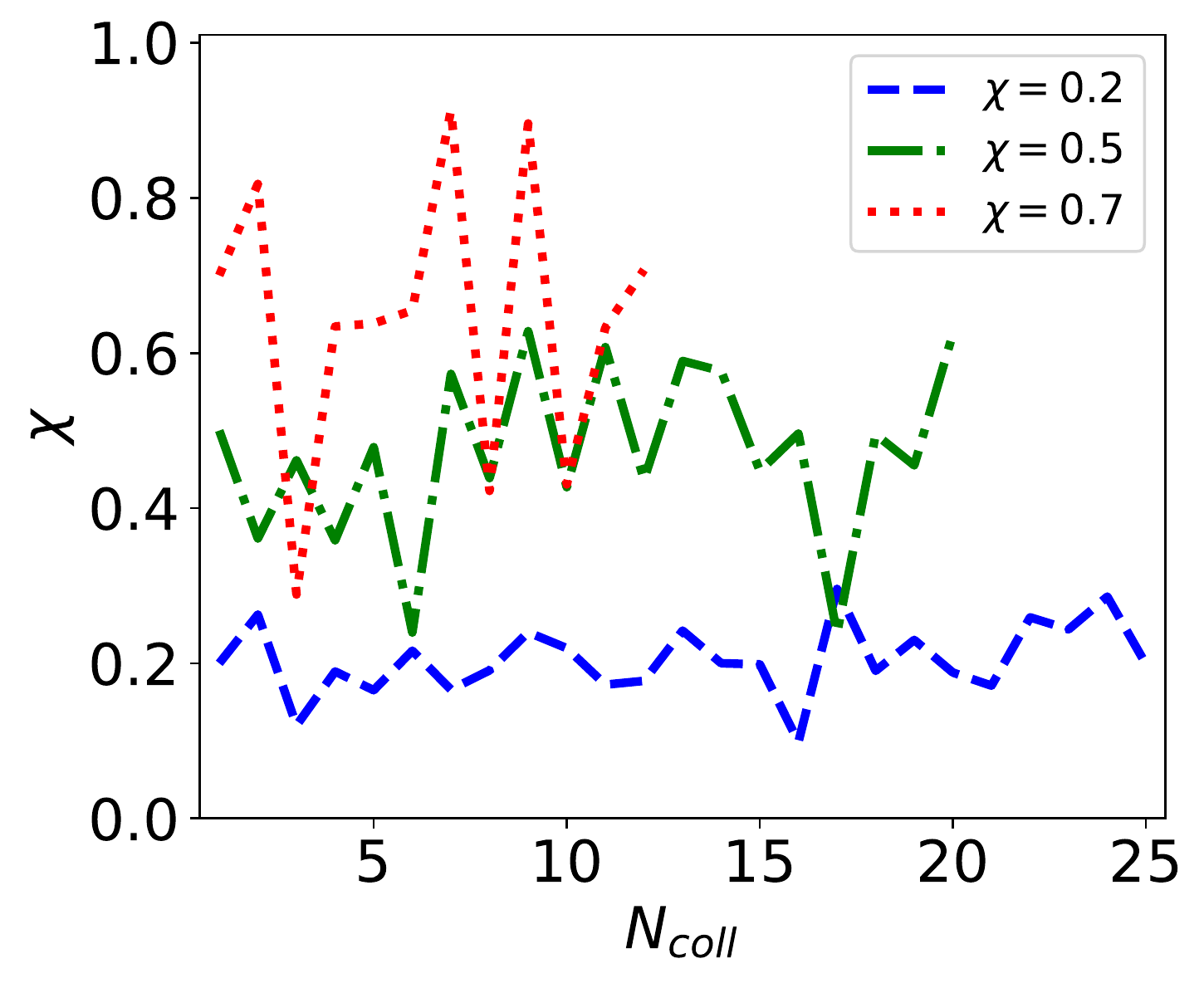}
\includegraphics[scale=0.55]{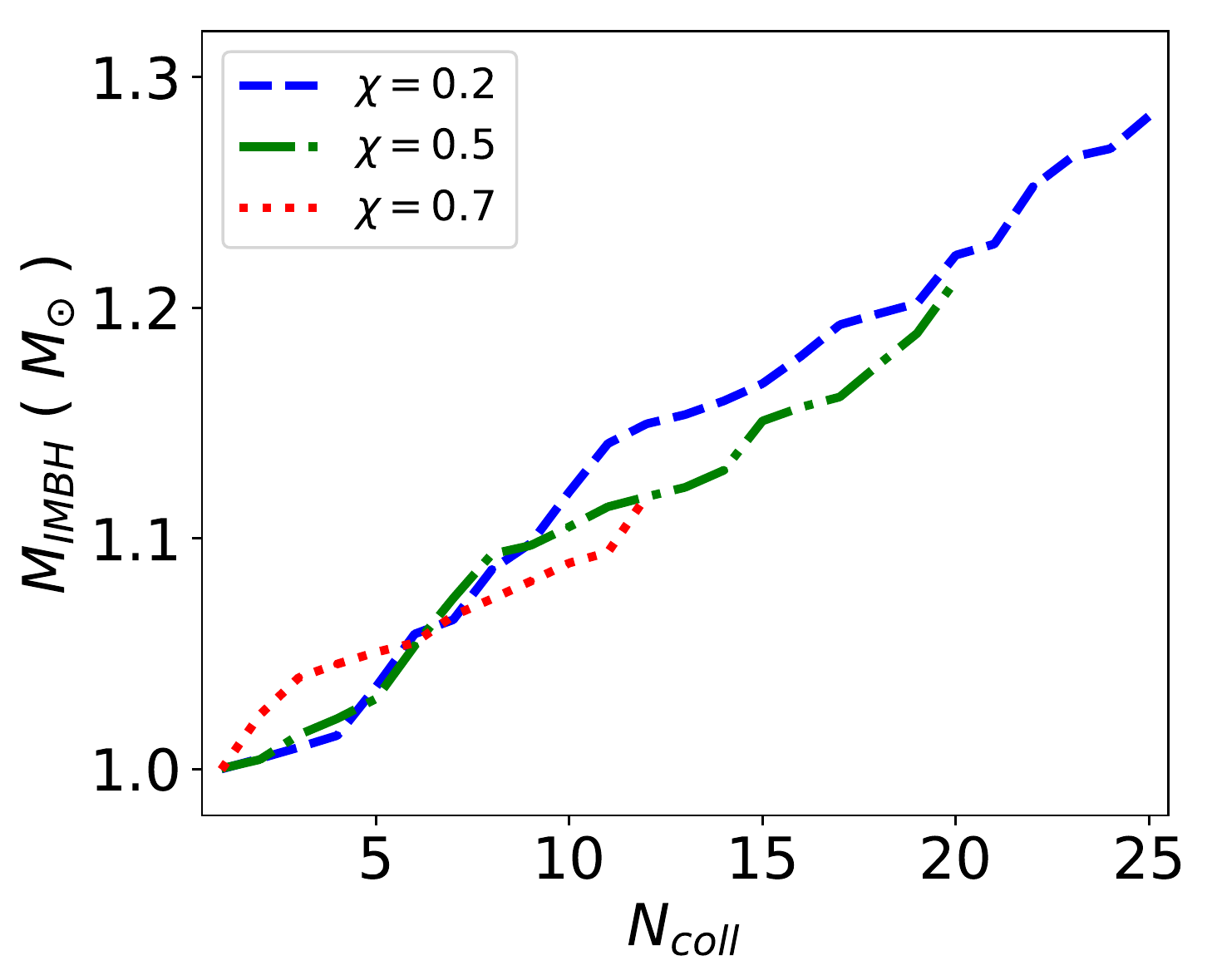}
\caption{Spin and mass evolution of an IMBH of $1.4\times 10^3\ \mathrm{M}_{\odot}$ for different initial spins of the IMBH and SBHs.}
\label{fig:evol_spin}
\end{figure}

As an example, Fig. \ref{fig:evol_spin} shows the spin and mass evolution of an IMBH of initial mass $1.4\times 10^3\ \mathrm{M}_{\odot}$ (hosted by a cluster of initial mass $1.4\times 10^5\ \mathrm{M}_{\odot}$) for different initial spins. As discussed, larger spins imply larger recoil velocities (see Eq. \ref{eqn:vkick}). As a consequence, the total number of collisions for $\chi=0.2$, $0.5$, and $0.7$ are $25$, $20$, and $12$, respectively. The spin of the IMBH random walks due to repeated mergers starting from the initial value
 (see Eq. \ref{eqn:finspin}), and stops at its final value when the IMBH is ejected from the cluster or has undergone $N_{\mathrm{coll}}$ merger events. For what concerns the mass, IMBHs with initial $\chi=0.2$ grow to a larger mass since they undergo a larger number of collisions with respect to the cases with initial $\chi=0.5$ and $\chi=0.7$.

\subsection{Number and rate of mergers and eccentricity effects}

As discussed, the eccentricity of the IMBH-SBH binary may have an important role since GW emission may drive the binary towards merger for eccentric systems. Equation (\ref{eqn:vkick}) also depends on the eccentricity through the correction $(1+e)$, which is valid only for small eccentricities \citep{hol08}, since the exact form of the kick velocity when $e$ approaches high values is not well known. As the eccentricity increases, GW emission may dominate the evolution of binary semi-major axis and eccentricity over dynamical interactions which leads to the circularization of the IMBH-BH binary, with eccentricities down to $\approx 10^{-4}$ until the merger due to GW emission \citep{pet64,ole06,ole09}. However, many SBHs may undergo mergers with the central IMBH through exchange with lower mass BHs and the resulting eccentricity can be high \citep{mil02b,gul06,hol08}. We run Model 4, in which all the merger events have eccentricity $0.2$ and fix the other parameters (see Tab. \ref{tab:models}). According to Eq. (\ref{eqn:vkick}), the kick velocities are $1.2\times$ larger than for circular inspiraling binaries. We find that the eccentricity does not change the overall results presented in the previous sections significantly.

The other two parameters that have to be taken into account are the average number of IMBH-SBH collisions $N_{\mathrm{coll}}$ and the average time $t_{\mathrm{coll}}$ between two subsequent collisions. For instance, if we assume that $t_{\mathrm{coll}}=50$ Myr, and the binary is made up of $1000\ \mathrm{M}_{\odot}$ IMBH and $10\ \mathrm{M}_{\odot}$ SBH orbiting in a circular orbit, the maximum semi-major axis to make the binary merge is $a_{\mathrm{GW}}\approx 0.2$ AU (see Eq. \ref{eqn:semgw}). The typical time to interact with a third body is $\propto a^{-1}_{\mathrm{GW}}$ and, in our case, $\approx 10^6$ yr \citep{ant16}. Such interactions remove a fraction $\approx 0.2 M_{SBH,3}/M_{\mathrm{IMBH}}+M_{\mathrm{SBH}}$ of the binary energy ($M_{SBH,3}$ is the mass of the third SBH), making the binary shrink even more. If we require a larger $t_{\mathrm{coll}}$, $a_{\mathrm{GW}}$ will be larger and the typical time to interact with a third body smaller, making interactions of the IMBH-SBH binary with surrounding BHs more important.

In general, $t_{\mathrm{coll}}$ is hard to define since it may depend on the cluster mass, and the total SBH depletion time may range from a few $100$ Myr to several Gyr depending on the cluster mass, as showed by $N$-body simulations of clusters with a central IMBH by \citet{lei14}. Recently, \citet{bre13} and \citet{mor13,mor15} show that globular clusters could retain a lot of BHs until present day.

In Model 5, we run models with $t_{\mathrm{coll}}=50$, $100$, $150$, $200$ Myr, respectively, to study the dependence of the results on this parameter. Such models correspond to a maximum number of IMBH-SBH merger events of $N_{\mathrm{coll}}=230$, $115$, $76$, $57$, respectively. The main effect of varying $t_{\mathrm{coll}}$ is on the inferred rate of IMRI events, as discussed in the next section. In the initial part of the IMBH-SBH merger history, a lot of mergers are due to low-mass GCs, from which the IMBH is ejected as a consequence of the recoil velocity after one or a few merger events. The final part is dominated by mergers with IMBHs in more massive GCs. Only a few of them survive and are not destroyed, and their IMBH undergo all $N_{\mathrm{coll}}$ events. For the others, since the merger events take place later in time for longer $t_{\mathrm{coll}}$, their mass is smaller because of tidal stripping by the Galactic field. As a consequence, or the GC is disrupted before undergoing $N_{\mathrm{coll}}$ merger events or its escape velocity is reduced and the IMBH may be ejected because of GW recoil velocity. Hence, the total number of merger events is larger for smaller $t_{\mathrm{coll}}$, and it is $\approx 30$\% smaller in the case $t_{\mathrm{coll}}=200$ Myr with respect to the case $t_{\mathrm{coll}}=50$ Myr. The other quantities are not significantly affected by the choice of $t_{\mathrm{coll}}$.

\section{Rate of IMRIs}
\label{sect:imris}

\begin{figure*} 
\centering
\begin{minipage}{18cm}
\includegraphics[scale=0.55]{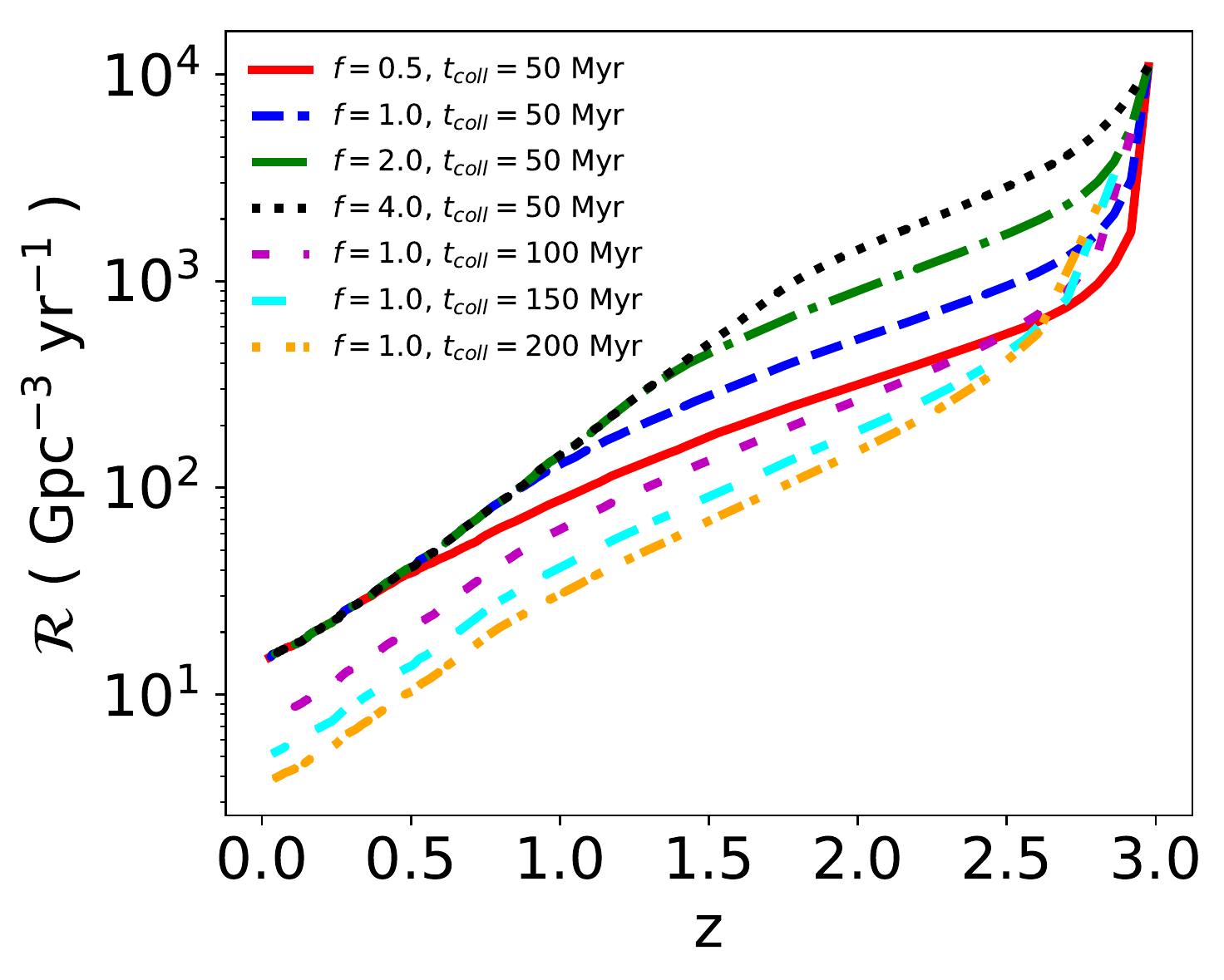}
\hspace{1cm}
\includegraphics[scale=0.55]{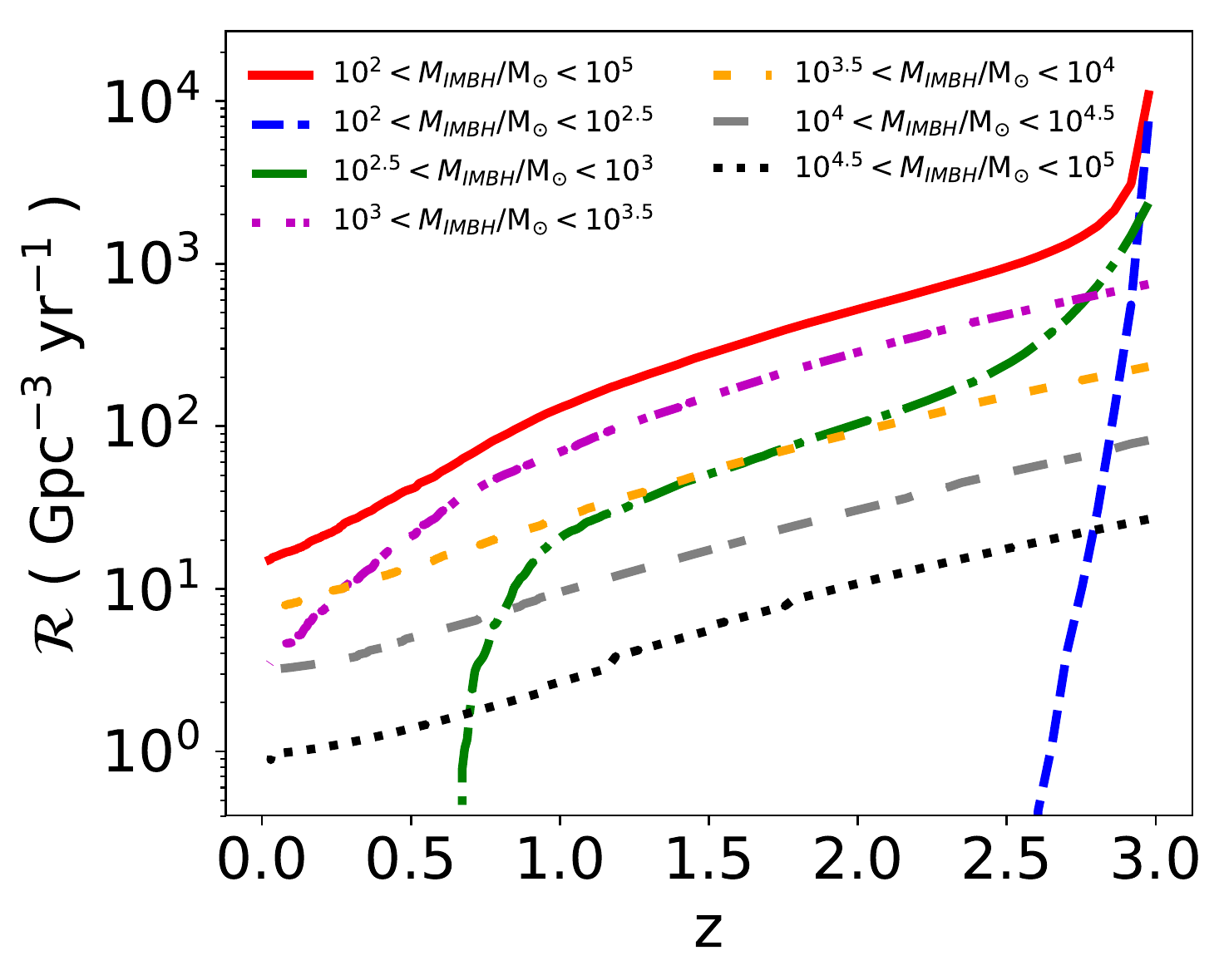}
\caption{Left panel: rate of IMRIs as a function of the redshift for Model 1 and Model 5. Right panel: rate of IMRIs as a function of the redshift for different IMBH masses for Model 1, when the fraction of the GC mass in IMBH is $f=1$\%. Note that LIGO upper limits are currently available only between $100$ and $300\,\mathrm{M}_{\odot}$ for $z$ close to zero where the rates are negligible (cf. dashed blue curve). However, at design sensitivity, LIGO/VIRGO/KAGRA will measure mergers with masses $300\,\mathrm{M}_{\odot}$--$3000\,\mathrm{M}_{\odot}$ (green and magenta dash-dotted curves), and ET and LISA may access nearly all IMBH mergers shown.}
\label{fig:rate_fimbh}
\end{minipage}
\end{figure*}

The study of IMBHs as sources of detectable GWs has been under scrutiny for years \citep{gul04,wil04,gul06,kon13}. In the inspiral phase, the angular-averaged characteristic dimensionless strain amplitude\footnote{i.e. the GW strain in a logarithmic frequency bin, or equivalently, the Fourier transform of the time-domain dimensionless GW strain multiplied by the frequency  \citep[cf. detector sensitivity,][]{moo15}.  Angular averaging is over the binary orientation, sky position, and the detector's antenna pattern. If the observation time is much less than the inspiral timescale and the source is circular, the source is approximately monochromatic with frequency $f_{\mathrm{GW}}\pm \frac12 T_{\mathrm{obs}}^{-1}$.} of the GWs emitted by a source at luminosity distance $D$ is \citep[e.g.][]{koc11}
\begin{align}\label{e:hc}
h_{\mathrm{c}}&\approx  
\begin{dcases}
\frac{4}{5} \frac{\mathrm{G}^{5/3}}{\mathrm{c}^{4}}
\frac{M_{\mathrm{IMBH}}^{5/3}M_{\mathrm{SBH}} f_{\mathrm{GW}}^{7/6}}{D  }T_{\mathrm{obs}}^{1/2}
& {\rm if}~~f_{\mathrm{GW}}\leq f_{\mathrm{crit}}\,,\nonumber\\
\frac{1}{\sqrt{30}\pi^{2/3}}\frac{\mathrm{G}^{5/6}}{\mathrm{c}^{3/2}}\frac{M_{\mathrm{IMBH}}^{1/3} M_{\mathrm{SBH}}^{1/2} }{D f_{\mathrm{GW}}^{1/6}} & {\rm if}~~f_{\mathrm{GW}}\geq f_{\mathrm{crit}}\,,
\end{dcases}
 \nonumber\\&=  
\begin{dcases}
6.2\times10^{-22} M_{\mathrm{IMBH,10^3M_{\odot}}}^{5/3} M_{\mathrm{SBH,10M_{\odot}}} 
f_{\mathrm{GW,10mHz}}^{\,5/6}T_{\mathrm{obs,yr}}^{1/2}
D_{\mathrm{Gpc}}^{-1}\\
\quad {\rm if}~~f_{\mathrm{GW}}\leq f_{\mathrm{crit}}\,,\\
1.4\times10^{-21} M_{\mathrm{IMBH,10^3M_{\odot}}}^{1/3} M_{\mathrm{SBH,10M_{\odot}}}^{1/2} 
f_{\mathrm{GW,10Hz}}^{\,-1/6}
D_{\mathrm{Gpc}}^{-1}\\
\quad {\rm if}~~f_{\mathrm{GW}}\geq f_{\mathrm{crit}}\,. 
\end{dcases}
\end{align}
where $f_{\rm crit}$ is given by the observation time $T_{\rm obs}$ as
\begin{equation}
f_{\rm crit} = 0.045\,M_{\mathrm{IMBH,10^3M_{\odot}}}^{-1/4}
M_{\mathrm{SBH,10M_{\odot}}}^{-3/8}T_{\mathrm{obs,yr}}^{-3/8}\,\mathrm{Hz}
\end{equation}
To save space, we have introduced a notation for quantities $x_a$ expressed with physical units $u$ as $x_{\mathrm{a},u} \equiv x_{\mathrm{a}}/u$, so that $x_{\mathrm{a},u}$ is dimensionless. For $f_{\mathrm{GW}}\lesssim f_{\rm crit}$ the GW frequency emitted by the binary is approximately constant during the observation time $T_{\mathrm{obs}}$ and for $f_{\mathrm{GW}}\gtrsim f_{\rm crit}$ the binary inspirals during the observation and spans a frequency range up to the innermost stable circular orbit. \textit{LISA} and aLIGO are expected to be sensitive to GW frequencies between $\approx 0.1$--$100$ mHz and $\approx 10$--$1000$ Hz, respectively. As a consequence, the GWs of IMRIs are potentially observable in these frequency ranges. During the inspiral, the GW frequency increases as the IMBH-SBH orbit shrinks until the last stable orbit which is followed by a rapid coalescence. Because of the quadrupolar nature of GWs, the typical frequency for circular binaries is twice the orbital frequency. At the innermost stable circular orbit (ISCO)
\begin{equation}
f_{\mathrm{GW,ISCO}}=2f_{\mathrm{orb}}\approx 4.4 M_{\mathrm{IMBH,10^3M_{\odot}}}^{-1}\mathrm{Hz}\ ,
\end{equation}
for non-spinning IMBHs and $f_{\mathrm{GW,ISCO}}$ is a factor $15$ higher for maximally spinning IMBHs.
The GW frequency before the binary reaches the ISCO is smaller than $f_{\mathrm{GW,ISCO}}$. After the final inspiral phase, the merger and ring-down phases emit GWs at a higher frequency with a characteristic ring-down frequency for zero spins is
\begin{equation}
f_{\mathrm{RD}}\approx 12 M_{\mathrm{IMBH,10^3M_{\odot}}}^{-1}\mathrm{Hz}\ .
\end{equation}
and it is a factor $\sim 10$ higher for nearly maximal spins \citep{ber09}. IMRIs involving IMBHs of a few hundred solar masses will first be observable by \textit{LISA} and then by LIGO. On the other hand, IMRIs involving IMBHs more massive than a few thousands solar masses will merge in the \textit{LISA} band and will not be detectable with LIGO. 

We now use the results from our simulated GC models to make predictions on the merger rate of IMRIs. The IMRI coalescence rate is still highly uncertain. The main limitation is to model consistently the evolution of GCs, where most of them are thought to take place, by also including GW energy losses. Moreover, clusters lose mass and inspiral towards the center of their galaxy across cosmic time. A conservative estimate of the merger rate of IMRIs formed in GCs is obtained by assuming that GCs in other galaxies have characteristics and histories similar to those in the Milky Way. We compute the merger rate between IMBHs and SBHs per unit volume and time as a function of the redshift $z$ as
\begin{equation}
\mathcal{R}(z)=n_{\mathrm{GC,total}}(z)\ \Gamma_{\mathrm{IMRI}}(z)\ ,
\label{eqn:rategw}
\end{equation}
where $n_{\mathrm{GC,total}}(z)=0.77\kappa [H(z)/H_0]^3$ Mpc$^{-3}$ is the spatial density of GCs and their remnants, including those which dissolved and those which survived by $z=0$, and $H(z)$ is the Hubble constant at redshift $z$\footnote{To compute $H(z)$ we use $\Omega_{m}=0.286$ and $\Omega_\Lambda=0.714$ for the matter and cosmological constant normalized density, respectively, and $H_0=70$ km s$^{-1}$ Mpc$^{-1}$ \citep{planck15}} and $\Gamma_{\mathrm{IMRI}}(z)$ is the redshifted\footnote{We take into account that the observed rate is redshifted by a factor $1/(1+z)$, where $z$ is the cosmological redshift} IMRI rate from a single GC which originally formed at redshift $z=3$. We choose $n_{\mathrm{GC,sur}}=0.77$ Mpc$^{-3}$ for the local GC density \citep{rod15}\footnote{A less conservative estimate by \citet{por00} predicts $n_{\mathrm{GC}}=2.9$ Mpc$^{-3}$ for the local GC density}. Moreover, we correct for $\kappa=N_{\rm GC,in}/N_{\rm GC,sur}=35$, where $N_{\rm GC,in}$ and $N_{\rm GC,sur}$ are the numbers of primordial GCs and GCs survived till present, respectively \citep{gne14}, to take into account IMRIs which happened in GCs that have dissolved by $z=0$. Finally, we compute $\Gamma_{\mathrm{IMRI}}$ from the results of our simulations (see Tab. \ref{tab:res_fimbh}).

\textit{LISA} may observe an IMBH-SBH binary in the inspiral phase and Advanced LIGO and VIRGO may detect the IMBH-SBH binary in the late inspiral and merger/ringdown phase at an SNR of 10 to a distance approximately given by\footnote{Here distance refers to luminosity distance and masses are redshifted masses, which are the source-frame mass times $(1+z)$.}  \citep{fla98,mil02a,gai11} 
\begin{align}
 D_{\mathrm{LIGO,inspiral}}&= 0.2\, 
 \left(\frac{M_{\mathrm{IMBH}}}{1000\mathrm{M}_{\odot}}\right)^{-1}
 \left(\frac{M_{\mathrm{SBH}}}{10\mathrm{M}_{\odot}}\right)^{1/2}\,\mathrm{Gpc}\,,\\
 D_{\mathrm{LIGO,RD}}&= 0.6\, 
 \left(\frac{M_{\mathrm{IMBH}}}{1000\mathrm{M}_{\odot}}\right)^{-3/2}
 \left(\frac{M_{\mathrm{SBH}}}{10\mathrm{M}_{\odot}}\right)\,\mathrm{Gpc}\,,\\
 D_{\mathrm{LISA,inspiral}}&= 0.6\, 
 \left(\frac{M_{\mathrm{IMBH}}}{1000\mathrm{M}_{\odot}}\right)^{1/2}
 \left(\frac{M_{\mathrm{SBH}}}{10\mathrm{M}_{\odot}}\right)^{1/2}\,\mathrm{Gpc}\,,\\
  D_{\mathrm{ET,inspiral}}&= 3.3\, 
 \left(\frac{M_{\mathrm{IMBH}}}{1000\mathrm{M}_{\odot}}\right)^{1/2}
 \left(\frac{M_{\mathrm{SBH}}}{10\mathrm{M}_{\odot}}\right)^{1/2}\,\mathrm{Gpc}\,.
\end{align}
The detection range of the VIRGO\footnote{http://www.virgo-gw.eu/}, KAGRA\footnote{http://gwcenter.icrr.u-tokyo.ac.jp}, and the two Advanced LIGO instruments is a factor 2 larger than that of a single interferometer given above.

Figure \ref{fig:rate_fimbh} shows the IMRI merger rate as a function of the redshift for Model 1 and Model 5 (left panel) from 2 Gyr until present. The right panel shows the merger rates for different IMBH masses for Model 1 for $f=1$\%. We divide our IMRIs in different mass bins since different instruments are expected to observe IMRIs with high sensitivity for different IMBH-SBH binary masses, as discussed. The largest rate comes for $10^3\ \mathrm{M}_{\odot}<M_{\mathrm{IMBH}}\le 10^4\ \mathrm{M}_{\odot}$, which will be detectable by either \textit{LISA} or ET. Assuming ET will be able to observe GW events with good $S/N$ up to $z\approx 2$ \citep{ama10}, our results predict a detection rate of $\approx 100-300$ Gpc$^{-3}$ yr$^{-1}$. Advanced LIGO, VIRGO, and KAGRA will be able to probe the low-end ($\lesssim 10^3\ \mathrm{M}_{\odot}$) of the IMBH population up to $z\sim 1.0$. Our models predict a rate of $0.5-20$ Gpc$^{-3}$ yr$^{-1}$ for $0.6\lesssim z\lesssim 1$ for $300 \lesssim M_{\mathrm{IMBH}}\lesssim 1000\ \mathrm{M}_{\odot}$. Lighter IMBHs are efficiently ejected by GW recoils or are in clusters dissolved by the galactic tidal field at $z\gtrsim 2.5$. \textit{LISA} may detect IMBHs of all masses, including the population of $\gtrsim 10^4\ \mathrm{M}_{\odot}$ IMBHs that are difficult to detect with even third generation Earth-based instruments like ET. The corresponding rate density of those massive IMBH-SBH mergers is $\approx 3-10$ Gpc$^{-3}$ yr$^{-1}$ at low redshift and it is above $\approx 100$ Gpc$^{-3}$ for $z\gtrsim 1.5$.

Our conclusions are consistent with the order of magnitude estimate of \citet{has16}. During their first observing run, LIGO did not detect GWs from IMBHBs \citep{abb16}. Recently, \citet{abb17} used such a result to constrain the rate of IMBH mergers with different masses and configurations. They found that for IMBHs of masses between $100\ \mathrm{M}_{\odot}$ and $300\ \mathrm{M}_{\odot}$ (with spins aligned with the binary orbital angular momentum) the merger rate is constrained to be $\lesssim 0.93$ Gpc$^{-3}$ yr$^{-1}$. The predicted merger rate in this mass range in our models is also very low for $z\leq 2.5$ (see blue dashed line in Figure \ref{fig:rate_fimbh}). However we emphasize that at design sensitivity Advanced LIGO/VIRGO/KAGRA may access IMBH masses between $300$ and $1000\,\mathrm{M}_{\odot}$ for which our models predict frequent mergers above $z\geq 0.6$ (see green dashed-dotted line in right panel). These upgraded instruments may also detect mergers for rapidly spinning IMBHs between $1000\,\mathrm{M}_{\odot}$ and $3000\,\mathrm{M}_{\odot}$, where the rates are much higher even at low redshift (see magenta dot-dash-dot curve in right panel).

As discussed in Section \ref{sect:res}, the parameters that affect the rate of IMRIs are the fraction of GC mass in IMBH $f$, the spin of the BHs $\chi$ and the total number of collisions $N_{\mathrm{coll}}$. Larger spins imply smaller inferred rates since the GW recoil kick can be as large as a few thousands km s$^{-1}$, while larger $f$'s and smaller $t_{\mathrm{coll}}$'s leads to larger rates since more massive IMBHs suffer from less intense recoil velocities and undergo a larger number of mergers, respectively. By measuring the mass, spin, and redshift distribution of IMBH mergers, GW observations may constrain the GC models.

\section{Conclusions}
\label{sect:conclusions}

The existence of IMBHs with masses $100\ \mathrm{M}_{\odot}\lesssim M\lesssim 10^5\ \mathrm{M}_{\odot}$ has not yet been confirmed directly. The formation of IMBHs is however theoretically very plausible, and one of the formation scenarios requires very dense environments as in the centres of GCs. The extrapolation of the observed correlation between SMBHs and their stellar environments to this mass range implies that these IMBHs could be present in GCs or in dwarf galaxies \citep{mer13,bau17,kiz17}. An IMBH would remain dark if not emitting because of accretion, but may influence the dynamical evolution of the GC. 

Gravitational waves will help in the hunt for the first direct evidence of IMBHs. IMBH-SBH binaries may form in GCs and represent so-called IMRIs, a down-scaled version of EMRIs. LISA will be able to detect tens of IMRIs at any given time \citep{mil02b,ama10,aep10}. As a consequence of the GW emission, the merger product will be imparted a gravitational wave recoil which, according to the mass-ratio and spins of the BHs, may be several thousands of km s$^{-1}$ times the square of the symmetric mass ratio \citep{lou11}. Due to the moderate mass-ratio of an IMRI and the shallow potential well of a GC, such recoils could be large enough to expel the IMBH from the cluster \citet{hol08}.

In this paper, we have investigated the possibility that primordial GCs were born with a central IMBH, which undergo merger events with SBHs in the cluster core \citep{por00,lei14,has16}. By means of a semi-analytical method, we have followed the evolution of the primordial cluster population in the galactic potential and the mergers of the binary IMBH-SBH systems \citep{gne14,fao17}. We have shown that low-mass IMBHs are usually ejected as a consequence of the GW recoil velocity, while massive IMBH ($\gtrsim 10^3\ \mathrm{M}_{\odot}$) are left in the Galactic field when their host GC is dissolved. In particular, we predict approximately $10^3$ ``bare'' IMBHs with mass $\gtrsim 10^3 \mathrm{M}_{\odot}$ each to be left without a host stellar cluster in the inner kpc of the galaxy (Figure~\ref{fig:fimbh_number}). The dynamical evolution in the galaxy of this population of IMBH has been studied by means of dynamical friction, but a more accurate study is needed to determine their final fate and distinguish such a population from the one predicted by cosmological models. Only a small fraction of the IMBHs remain in their host cluster without being ejected due to repeated GW recoil kicks and without the host GC being tidally disrupted by the Galactic field. We have illustrated that the typical chirp mass and mass ratio of an IMRI depend on the initial fraction $f$ of GC mass in IMBH and also on the slope $\zeta$ of the SBHs distribution. Larger $f$ and/or shallower slope $\zeta$ imply larger typical chirp masses of the IMBH-SBH merger event. We have also investigated the role of the spin in an IMRI. We found that the spin of the final merger product is mostly determined by the initial spin of the IMBH, with a spread due to the geometry of the angular momentum and of the spins at the moment of the merging. Moreover, we have shown that larger spins imply larger GW recoil velocity, which lead to $\approx 7$\% of the primordial IMBHs to be ejected from the galaxy. This amounts to a total of $\approx 500$ IMBHs ejected per Milky-Way type galaxy and implies an intergalactic IMBH number density of 
\begin{equation}
n_{\mathrm{intergalactic\, IMBH}} \sim 10\, \mathrm{Mpc}^{-3}\,.
\end{equation}
 These estimates are not sensitive to the IMRI eccentricity. However, if the IMBH-SBH collision rate at early times is higher than in our fiducial model, i.e. $1/(50\,{\rm Myr})$, then the number of IMRI GW events increases.

We have used the average number of IMBH-SBH merger events in our models to predict the rate of IMRIs for current and upcoming GW detections. While 
IMBHs of any mass can be potentially observed by \textit{LISA}, LIGO is able to spot only IMBHs with masses of a few hundred $\mathrm{M}_{\odot}$ if it is non-spinning and a few thousand $\mathrm{M}_{\odot}$ if it is maximally spinning. The IMBH-SBH merger rate density decreases in time, from a value $\mathcal{R}\approx 100$ Gpc$^{-3}$ yr$^{-1}$ at $z\gtrsim 2$ to $\mathcal{R}\approx 0.5$ Gpc$^{-3}$ yr$^{-1}$ at low redshift (Figure~\ref{fig:rate_fimbh}). The rate density is varies greatly with the IMBH mass, it is dominated by IMBH masses in the range $10^3\ \mathrm{M}_{\odot}\lesssim M_{\mathrm{IMBH}}\lesssim 10^4\ \mathrm{M}_{\odot}$. Current LIGO upper limits exist only for masses $M_{\mathrm{IMBH}}\leq 300$, where the predicted rate is negligible below $z\lesssim 2.6$. These lighter IMBHs are efficiently ejected by GW recoils or are in clusters dissolved by the galactic tidal field. However, we predict that LIGO/VIRGO/KAGRA may detect IMBH mergers at design sensitivity with $\mathcal{R}\approx 1-10$ Gpc$^{-3}$ yr$^{-1}$ if it can access the mass range between $300$--$1000\ \mathrm{M}_{\odot}$ for $z\gtrsim 0.6$ and $\mathcal{R}\approx 5-10$ Gpc$^{-3}$ for IMBHs with $1000$--$3000\ \mathrm{M}_{\odot}$ in the local universe (Figure~\ref{fig:rate_fimbh}).
Assuming \textit{LISA} will be able to observe GW events up to $z\approx 1$ and ET up to $z\approx 2$ \citep{ama10}, our results predict a detection rate density for $\approx 100-300$ Gpc$^{-3}$ yr$^{-1}$ for IMBHs with mass of $\approx 10^3$-$10^4\ \mathrm{M}_{\odot}$ with ET and \textit{LISA} and a detection rate density for $\approx 3-10$ Gpc$^{-3}$ yr$^{-1}$ from IMBH-SBH binaries with masses $\gtrsim 10^4\ \mathrm{M}_{\odot}$ observable with \textit{LISA}. 

We note that the rates at low redshifts may be significantly higher if young massive star clusters host IMBHs. In addition to generating distinct GW events, the large number of inspirals of IMBH-SBH systems at cosmological distances generates an unresolved diffuse GW background \citep{leo04} which may affect \textit{LISA} and ET observations.

By measuring the mass, spin, and redshift distribution of IMBH mergers, GW observations may help to improve our understanding of GC and galaxy evolution.

\section{Acknowledgements}

We thank Daniel D'Orazio, Oleg Gnedin, and Deirdre Shoemaker for useful discussions and comments. GF acknowledges hospitality from the E\"{o}tv\"{o}s Lor\'{a}nd University of Budapest. 
This project has received funding from the European Research Council (ERC) under the European Union's Horizon 2020 research and innovation programme under grant agreement No 638435 (GalNUC) and by the Hungarian National Research, Development, and Innovation Office grant NKFIH KH-125675 (to BK). This work was performed by BK in part at the Aspen Center for Physics, which is supported by National Science Foundation grant PHY-1607761. This work was performed in part (by BK) at the Aspen Center for Physics, which is supported by National Science Foundation grant PHY-1607761. IG was supported in part by Harvard University and the Institute for Theory and Computation. 

\bibliographystyle{yahapj}
\bibliography{refs}

\begin{thebibliography}{}
\providecommand\natexlab[1]{#1}
\providecommand\JournalTitle[1]{#1}

\bibitem[{{Abbott} {et~al.}(2016){Abbott}, {Abbott}, {Abbott}, {Abernathy},
  {Acernese}, {Ackley}, {Adams}, {Adams}, {Addesso}, {Adhikari}, \&
  et~al.}]{abb16}
{Abbott}, B.~P., {Abbott}, R., {Abbott}, T.~D., {et~al.} 2016,
  \href{http://dx.doi.org/10.3847/2041-8205/818/2/L22}{\JournalTitle{\apjl},
  818, L22}

\bibitem[{{Abbott} {et~al.}(2017){Abbott}, {Abbott}, {Abbott}, {Acernese},
  {Ackley}, {Adams}, {Adams}, {Addesso}, {Adhikari}, {Adya}, \& et~al.}]{abb17}
---. 2017,
  \href{http://dx.doi.org/10.1103/PhysRevD.96.022001}{\JournalTitle{\prd}, 96,
  022001}

\bibitem[{{Abramowicz} {et~al.}(2004){Abramowicz}, {Klu{\'z}niak},
  {McClintock}, \& {Remillard}}]{abr04}
{Abramowicz}, M.~A., {Klu{\'z}niak}, W., {McClintock}, J.~E., \& {Remillard},
  R.~A. 2004, \href{http://dx.doi.org/10.1086/422810}{\JournalTitle{\apjl},
  609, L63}

\bibitem[{{Amaro-Seoane} \& {Chen}(2016)}]{ama16}
{Amaro-Seoane}, P., \& {Chen}, X. 2016,
  \href{http://dx.doi.org/10.1093/mnras/stw503}{\JournalTitle{\mnras}, 458,
  3075}

\bibitem[{{Amaro-Seoane} {et~al.}(2010){Amaro-Seoane}, {Eichhorn}, {Porter}, \&
  {Spurzem}}]{aep10}
{Amaro-Seoane}, P., {Eichhorn}, C., {Porter}, E.~K., \& {Spurzem}, R. 2010,
  \href{http://dx.doi.org/10.1111/j.1365-2966.2009.15842.x}{\JournalTitle{\mnras},
  401, 2268}

\bibitem[{{Amaro-Seoane} {et~al.}(2007){Amaro-Seoane}, {Gair}, {Freitag},
  {Miller}, {Mandel}, {Cutler}, \& {Babak}}]{ama07}
{Amaro-Seoane}, P., {Gair}, J.~R., {Freitag}, M., {et~al.} 2007,
  \href{http://dx.doi.org/10.1088/0264-9381/24/17/R01}{\JournalTitle{Classical
  and Quantum Gravity}, 24, R113}

\bibitem[{{Amaro-Seoane} \& {Santamar{\'{\i}}a}(2010)}]{ama10}
{Amaro-Seoane}, P., \& {Santamar{\'{\i}}a}, L. 2010,
  \href{http://dx.doi.org/10.1088/0004-637X/722/2/1197}{\JournalTitle{\apj},
  722, 1197}

\bibitem[{{Antonini}(2013)}]{ant13}
{Antonini}, F. 2013,
  \href{http://dx.doi.org/10.1088/0004-637X/763/1/62}{\JournalTitle{\apj}, 763,
  62}

\bibitem[{{Antonini} \& {Rasio}(2016)}]{ant16}
{Antonini}, F., \& {Rasio}, F.~A. 2016,
  \href{http://dx.doi.org/10.3847/0004-637X/831/2/187}{\JournalTitle{\apj},
  831, 187}

\bibitem[{{Arca-Sedda}(2016)}]{arc16}
{Arca-Sedda}, M. 2016,
  \href{http://dx.doi.org/10.1093/mnras/stv2265}{\JournalTitle{\mnras}, 455,
  35}

\bibitem[{{Barack} \& {Cutler}(2004)}]{leo04}
{Barack}, L., \& {Cutler}, C. 2004,
  \href{http://dx.doi.org/10.1103/PhysRevD.70.122002}{\JournalTitle{Physical
  Review D}, 70, 122002}

\bibitem[{{Baumgardt}(2017)}]{bau17}
{Baumgardt}, H. 2017,
  \href{http://dx.doi.org/10.1093/mnras/stw2488}{\JournalTitle{\mnras}, 464,
  2174}

\bibitem[{{Belczynski} {et~al.}(2017){Belczynski}, {Klencki}, {Meynet},
  {Fryer}, {Brown}, {Chruslinska}, {Gladysz}, {O'Shaughnessy}, {Bulik},
  {Berti}, {Holz}, {Gerosa}, {Giersz}, {Ekstrom}, {Georgy}, {Askar}, \&
  {Lasota}}]{bel17}
{Belczynski}, K., {Klencki}, J., {Meynet}, G., {et~al.} 2017,
  \JournalTitle{ArXiv e-prints},
  \href{http://arxiv.org/abs/1706.07053}{{\sffamily arXiv:1706.07053
  [astro-ph.HE]}}

\bibitem[{{Berti} {et~al.}(2009){Berti}, {Cardoso}, \& {Starinets}}]{ber09}
{Berti}, E., {Cardoso}, V., \& {Starinets}, A.~O. 2009,
  \href{http://dx.doi.org/10.1088/0264-9381/26/16/163001}{\JournalTitle{Classical
  and Quantum Gravity}, 26, 163001}

\bibitem[{{Binney} \& {Tremaine}(2008)}]{bin08}
{Binney}, J., \& {Tremaine}, S. 2008, {Galactic Dynamics: Second Edition}
  (Princeton University Press)

\bibitem[{{Breen} \& {Heggie}(2013)}]{bre13}
{Breen}, F.~G., \& {Heggie}, D.~C. 2013,
  \href{http://dx.doi.org/10.1093/mnras/stt628}{\JournalTitle{\mnras}, 432,
  2779}

\bibitem[{{Buliga} {et~al.}(2011){Buliga}, {Globina}, {Gnedin},
  {Natsvlishvili}, {Piotrovich}, \& {Shakht}}]{bul11}
{Buliga}, S.~D., {Globina}, V.~I., {Gnedin}, Y.~N., {et~al.} 2011,
  \href{http://dx.doi.org/10.1007/s10511-011-9204-7}{\JournalTitle{Astrophysics},
  54, 548}

\bibitem[{{Buonanno} {et~al.}(2008){Buonanno}, {Kidder}, \& {Lehner}}]{buo08}
{Buonanno}, A., {Kidder}, L.~E., \& {Lehner}, L. 2008,
  \href{http://dx.doi.org/10.1103/PhysRevD.77.026004}{\JournalTitle{PHYSICAL
  REVIEW D}, 77, 026004}

\bibitem[{{Capuzzo-Dolcetta} \& {Fragione}(2015)}]{cap15}
{Capuzzo-Dolcetta}, R., \& {Fragione}, G. 2015,
  \href{http://dx.doi.org/10.1093/mnras/stv2123}{\JournalTitle{\mnras}, 454,
  2677}

\bibitem[{{Chernoff} \& {Weinberg}(1990)}]{che90}
{Chernoff}, D.~F., \& {Weinberg}, M.~D. 1990,
  \href{http://dx.doi.org/10.1086/168451}{\JournalTitle{\apj}, 351, 121}

\bibitem[{{Davis} {et~al.}(2011){Davis}, {Narayan}, {Zhu}, {Barret}, {Farrell},
  {Godet}, {Servillat}, \& {Webb}}]{dav11}
{Davis}, S.~W., {Narayan}, R., {Zhu}, Y., {et~al.} 2011,
  \href{http://dx.doi.org/10.1088/0004-637X/734/2/111}{\JournalTitle{\apj},
  734, 111}

\bibitem[{{Fabbiano}(2006)}]{fab06}
{Fabbiano}, G. 2006,
  \href{http://dx.doi.org/10.1146/annurev.astro.44.051905.092519}{\JournalTitle{\araa},
  44, 323}

\bibitem[{{Fishbach} {et~al.}(2017){Fishbach}, {Holz}, \& {Farr}}]{fis17}
{Fishbach}, M., {Holz}, D.~E., \& {Farr}, B. 2017,
  \href{http://dx.doi.org/10.3847/2041-8213/aa7045}{\JournalTitle{\apjl}, 840,
  L24}

\bibitem[{{Flanagan} \& {Hughes}(1998)}]{fla98}
{Flanagan}, {\'E}.~{\'E}., \& {Hughes}, S.~A. 1998,
  \href{http://dx.doi.org/10.1103/PhysRevD.57.4535}{\JournalTitle{\prd}, 57,
  4535}

\bibitem[{{Fragione} {et~al.}(2018){Fragione}, {Antonini}, \& {Gnedin}}]{fao17}
{Fragione}, G., {Antonini}, F., \& {Gnedin}, O.~Y. 2018,
  \href{http://dx.doi.org/10.1093/mnras/sty183}{\JournalTitle{ArXiv e-prints}},
  \href{http://arxiv.org/abs/1709.03534}{{\sffamily arXiv:1709.03534}}

\bibitem[{{Fragione} \& {Capuzzo-Dolcetta}(2016)}]{fra16}
{Fragione}, G., \& {Capuzzo-Dolcetta}, R. 2016,
  \href{http://dx.doi.org/10.1093/mnras/stw531}{\JournalTitle{\mnras}, 458,
  2596}

\bibitem[{{Fragione} {et~al.}(2017){Fragione}, {Capuzzo-Dolcetta}, \&
  {Kroupa}}]{fra17}
{Fragione}, G., {Capuzzo-Dolcetta}, R., \& {Kroupa}, P. 2017,
  \href{http://dx.doi.org/10.1093/mnras/stx106}{\JournalTitle{\mnras}, 467,
  451}

\bibitem[{{Fragione} \& {Ginsburg}(2017)}]{fgg17}
{Fragione}, G., \& {Ginsburg}, I. 2017,
  \href{http://dx.doi.org/10.1093/mnras/stw3213}{\JournalTitle{\mnras}, 466,
  1805}

\bibitem[{{Fragione} \& {Gualandris}(2018)}]{fgu18}
{Fragione}, G., \& {Gualandris}, A. 2018,
  \href{http://dx.doi.org/10.1093/mnras/sty145}{\JournalTitle{\mnras}},
  \href{http://arxiv.org/abs/1801.02588}{{\sffamily arXiv:1801.02588}}

\bibitem[{{Fragione} \& {Loeb}(2017)}]{frl17}
{Fragione}, G., \& {Loeb}, A. 2017,
  \href{http://dx.doi.org/10.1016/j.newast.2017.03.002}{\JournalTitle{New
  Astronomy}, 55, 32}

\bibitem[{{Fragione} \& {Sari}(2018)}]{frs18}
{Fragione}, G., \& {Sari}, R. 2018,
  \href{http://dx.doi.org/10.3847/1538-4357/aaa0d7}{\JournalTitle{\apj}, 852,
  51}

\bibitem[{{Freitag} {et~al.}(2006){Freitag}, {G{\"u}rkan}, \& {Rasio}}]{fre06}
{Freitag}, M., {G{\"u}rkan}, M.~A., \& {Rasio}, F.~A. 2006,
  \href{http://dx.doi.org/10.1111/j.1365-2966.2006.10096.x}{\JournalTitle{\mnras},
  368, 141}

\bibitem[{{Gair} {et~al.}(2011){Gair}, I., {Miller}, \& {Volonteri}}]{gai11}
{Gair}, J.~R., I., M., {Miller}, M.~C., \& {Volonteri}, M. 2011,
  \href{http://dx.doi.org/10.1007/s10714-010-1104-3}{\JournalTitle{General
  Relativity and Gravitation}, 43, 485}

\bibitem[{{Gieles} \& {Baumgardt}(2008)}]{gie08}
{Gieles}, M., \& {Baumgardt}, H. 2008,
  \href{http://dx.doi.org/10.1111/j.1745-3933.2008.00515.x}{\JournalTitle{\mnras},
  389, L28}

\bibitem[{{Gieles} {et~al.}(2011){Gieles}, {Heggie}, \& {Zhao}}]{gie11}
{Gieles}, M., {Heggie}, D.~C., \& {Zhao}, H. 2011,
  \href{http://dx.doi.org/10.1111/j.1365-2966.2011.18320.x}{\JournalTitle{\mnras},
  413, 2509}

\bibitem[{{Giersz} {et~al.}(2015){Giersz}, {Leigh}, {Hypki}, {L\"{u}tzgendorf},
  \& {Askar}}]{gie15}
{Giersz}, M., {Leigh}, N.~W., {Hypki}, A., {L\"{u}tzgendorf}, N., \& {Askar},
  A. 2015,
  \href{http://dx.doi.org/10.1093/mnras/stv2162}{\JournalTitle{\mnras}, 454,
  3150}

\bibitem[{{Gnedin} {et~al.}(2014){Gnedin}, {Ostriker}, \& {Tremaine}}]{gne14}
{Gnedin}, O.~Y., {Ostriker}, J.~P., \& {Tremaine}, S. 2014,
  \href{http://dx.doi.org/10.1088/0004-637X/785/1/71}{\JournalTitle{\apj}, 785,
  71}

\bibitem[{{Gonz{\'a}lez} {et~al.}(2007){Gonz{\'a}lez}, {Sperhake},
  {Br{\"u}gmann}, {Hannam}, \& {Husa}}]{gon07}
{Gonz{\'a}lez}, J.~A., {Sperhake}, U., {Br{\"u}gmann}, B., {Hannam}, M., \&
  {Husa}, S. 2007,
  \href{http://dx.doi.org/10.1103/PhysRevLett.98.091101}{\JournalTitle{Physical
  Review Letters}, 98, 091101}

\bibitem[{{Gualandris} {et~al.}(2010){Gualandris}, {Gillessen}, \&
  {Merritt}}]{gua10}
{Gualandris}, A., {Gillessen}, S., \& {Merritt}, D. 2010,
  \href{http://dx.doi.org/10.1111/j.1365-2966.2010.17373.x}{\JournalTitle{\mnras},
  409, 1146}

\bibitem[{{Gualandris} \& {Merritt}(2009)}]{gua09}
{Gualandris}, A., \& {Merritt}, D. 2009,
  \href{http://dx.doi.org/10.1088/0004-637X/705/1/361}{\JournalTitle{\apj},
  705, 361}

\bibitem[{{G{\"u}ltekin} {et~al.}(2004){G{\"u}ltekin}, {Miller}, \&
  {Hamilton}}]{gul04}
{G{\"u}ltekin}, K., {Miller}, M.~C., \& {Hamilton}, D.~P. 2004,
  \href{http://dx.doi.org/10.1086/424809}{\JournalTitle{\apj}, 616, 221}

\bibitem[{{G{\"u}ltekin} {et~al.}(2006){G{\"u}ltekin}, {Miller}, \&
  {Hamilton}}]{gul06}
---. 2006, \href{http://dx.doi.org/10.1086/499917}{\JournalTitle{\apj}, 640,
  156}

\bibitem[{{Haster} {et~al.}(2016){Haster}, {Antonini}, {Kalogera}, \&
  {Mandel}}]{has16}
{Haster}, C.-J., {Antonini}, F., {Kalogera}, V., \& {Mandel}, I. 2016,
  \href{http://dx.doi.org/10.3847/0004-637X/832/2/192}{\JournalTitle{\apj},
  832, 192}

\bibitem[{{Hofmann} {et~al.}(2016){Hofmann}, {Barausse}, \& {Rezzolla}}]{hof16}
{Hofmann}, F., {Barausse}, E., \& {Rezzolla}, L. 2016,
  \href{http://dx.doi.org/10.3847/2041-8205/825/2/L19}{\JournalTitle{\apj
  Lett.}, 625, L19}

\bibitem[{{Holley-Bockelmann} {et~al.}(2008){Holley-Bockelmann},
  {G{\"u}ltekin}, {Shoemaker}, \& {Yunes}}]{hol08}
{Holley-Bockelmann}, K., {G{\"u}ltekin}, K., {Shoemaker}, D., \& {Yunes}, N.
  2008, \href{http://dx.doi.org/10.1086/591218}{\JournalTitle{\apj}, 686, 829}

\bibitem[{{Hopman} \& {Alexander}(2006)}]{hop06}
{Hopman}, C., \& {Alexander}, T. 2006,
  \href{http://dx.doi.org/10.1086/506273}{\JournalTitle{\apjl}, 645, L133}

\bibitem[{{Hurley} {et~al.}(2000){Hurley}, {Pols}, \& {Tout}}]{hur00}
{Hurley}, J.~R., {Pols}, O.~R., \& {Tout}, C.~A. 2000,
  \href{http://dx.doi.org/10.1046/j.1365-8711.2000.03426.x}{\JournalTitle{\mnras},
  315, 543}

\bibitem[{{Jiang} {et~al.}(2008){Jiang}, {Jing}, {Faltenbacher}, {Lin}, \&
  {Li}}]{jia08}
{Jiang}, C.~Y., {Jing}, Y.~P., {Faltenbacher}, A., {Lin}, W.~P., \& {Li}, C.
  2008, \href{http://dx.doi.org/10.1086/526412}{\JournalTitle{\apj}, 675, 1095}

\bibitem[{{Kaaret} {et~al.}(2017){Kaaret}, {Feng}, \& {Roberts}}]{kaa17}
{Kaaret}, P., {Feng}, H., \& {Roberts}, T.~P. 2017,
  \href{http://dx.doi.org/10.1146/annurev-astro-091916-055259}{\JournalTitle{\araa},
  55, 303}

\bibitem[{{K{\i}z{\i}ltan} {et~al.}(2017){K{\i}z{\i}ltan}, {Baumgardt}, \&
  {Loeb}}]{kiz17}
{K{\i}z{\i}ltan}, B., {Baumgardt}, H., \& {Loeb}, A. 2017,
  \href{http://dx.doi.org/10.1038/nature21361}{\JournalTitle{\nat}, 542, 203}

\bibitem[{{Kocsis} {et~al.}(2012){Kocsis}, {Ray}, \& {Portegies Zwart}}]{koc12}
{Kocsis}, B., {Ray}, A., \& {Portegies Zwart}, S. 2012,
  \href{http://dx.doi.org/10.1088/0004-637X/752/1/67}{\JournalTitle{\apj}, 752,
  67}

\bibitem[{{Kocsis} {et~al.}(2011){Kocsis}, {Yunes}, \& {Loeb}}]{koc11}
{Kocsis}, B., {Yunes}, N., \& {Loeb}, A. 2011,
  \href{http://dx.doi.org/10.1103/PhysRevD.84.024032}{\JournalTitle{\prd}, 84,
  024032}

\bibitem[{{Konstantinidis} {et~al.}(2013){Konstantinidis}, {Amaro-Seoane}, \&
  {Kokkotas}}]{kon13}
{Konstantinidis}, S., {Amaro-Seoane}, P., \& {Kokkotas}, K.~D. 2013,
  \href{http://dx.doi.org/10.1051/0004-6361/201219620}{\JournalTitle{\aap},
  557, A135}

\bibitem[{{Kozai}(1962)}]{koz62}
{Kozai}, Y. 1962, \href{http://dx.doi.org/10.1086/108790}{\JournalTitle{\aj},
  67, 591}

\bibitem[{{Kroupa}(2001)}]{kro01}
{Kroupa}, P. 2001,
  \href{http://dx.doi.org/10.1046/j.1365-8711.2001.04022.x}{\JournalTitle{\mnras},
  322, 231}

\bibitem[{{Kruijssen} \& {L\"{u}tzgendorf}(2013)}]{krl13}
{Kruijssen}, J.~M.~D., \& {L\"{u}tzgendorf}, N. 2013,
  \href{http://dx.doi.org/10.1093/mnrasl/slt073}{\JournalTitle{\mnras}, 434,
  L41}

\bibitem[{{Kushnir} {et~al.}(2016){Kushnir}, {Zaldarriaga}, {Kollmeier}, \&
  {Waldman}}]{kus16}
{Kushnir}, D., {Zaldarriaga}, M., {Kollmeier}, J.~A., \& {Waldman}, R. 2016,
  \href{http://dx.doi.org/10.1093/mnras/stw1684}{\JournalTitle{\mnras}, 462,
  844}

\bibitem[{{Leigh} {et~al.}(2013){Leigh}, {B\"{o}ker}, {Maccarone}, \&
  {Perets}}]{lee13}
{Leigh}, N.~W.~C., {B\"{o}ker}, T., {Maccarone}, T.~J., \& {Perets}, H.~B.
  2013, \href{http://dx.doi.org/10.1093/mnras/sts554}{\JournalTitle{\mnras},
  429, 2997}

\bibitem[{{Leigh} {et~al.}(2014){Leigh}, {L{\"u}tzgendorf}, {Geller},
  {Maccarone}, {Heinke}, \& {Sesana}}]{lei14}
{Leigh}, N.~W.~C., {L{\"u}tzgendorf}, N., {Geller}, A.~M., {et~al.} 2014,
  \href{http://dx.doi.org/10.1093/mnras/stu1437}{\JournalTitle{\mnras}, 444,
  29}

\bibitem[{{Leigh} \& {Sills}(2011)}]{lei11}
{Leigh}, N.~W.~C., \& {Sills}, A. 2011,
  \href{http://dx.doi.org/10.1111/j.1365-2966.2010.17609.x}{\JournalTitle{\mnras},
  410, 2370}

\bibitem[{{Lidov}(1962)}]{lid62}
{Lidov}, M.~L. 1962,
  \href{http://dx.doi.org/10.1016/0032-0633(62)90129-0}{\JournalTitle{\planss},
  9, 719}

\bibitem[{{Lousto} {et~al.}(2010){Lousto}, {Campanelli}, {Zlochower}, \&
  {Nakano}}]{lou10}
{Lousto}, C.~O., {Campanelli}, M., {Zlochower}, Y., \& {Nakano}, H. 2010,
  \href{http://dx.doi.org/10.1088/0264-9381/27/11/114006}{\JournalTitle{Classical
  and Quantum Gravity}, 27, 114006}

\bibitem[{{Lousto} \& {Zlochower}(2008)}]{lou08}
{Lousto}, C.~O., \& {Zlochower}, Y. 2008,
  \href{http://dx.doi.org/10.1103/PhysRevD.77.044028}{\JournalTitle{\prd}, 77,
  044028}

\bibitem[{{Lousto} \& {Zlochower}(2011)}]{lou11}
---. 2011,
  \href{http://dx.doi.org/10.1103/PhysRevLett.107.231102}{\JournalTitle{Physical
  Review Letters}, 107, 231102}

\bibitem[{{Lousto} {et~al.}(2012){Lousto}, {Zlochower}, {Dotti}, \&
  {Volonteri}}]{lou12}
{Lousto}, C.~O., {Zlochower}, Y., {Dotti}, M., \& {Volonteri}, M. 2012,
  \href{http://dx.doi.org/10.1103/PhysRevD.85.084015}{\JournalTitle{\prd}, 85,
  084015}

\bibitem[{{L{\"u}tzgendorf} {et~al.}(2013){L{\"u}tzgendorf}, {Baumgardt}, \&
  {Kruijssen}}]{lut13}
{L{\"u}tzgendorf}, N., {Baumgardt}, H., \& {Kruijssen}, J.~M.~D. 2013,
  \href{http://dx.doi.org/10.1051/0004-6361/201321927}{\JournalTitle{\aap},
  558, A117}

\bibitem[{{L\"{u}tzgendorf} {et~al.}(2012){L\"{u}tzgendorf}, {Kissler-Patig},
  {Gebhardt}, {Baumgardt}, {Noyola}, {Jalali}, {de Zeeuw}, \&
  {Neumayer}}]{lkg13}
{L\"{u}tzgendorf}, N., {Kissler-Patig}, M., {Gebhardt}, K., {et~al.} 2012,
  \href{http://dx.doi.org/10.1051/0004-6361/201219375}{\JournalTitle{\aap},
  542, A129}

\bibitem[{{L\"{u}tzgendorf} {et~al.}(2013{\natexlab{a}}){L\"{u}tzgendorf},
  {Kissler-Patig}, {Gebhardt}, {Baumgardt}, {Noyola}, {de Zeeuw}, {Neumayer},
  {Jalali}, \& {Feldmeier}}]{luk13}
---. 2013{\natexlab{a}},
  \href{http://dx.doi.org/10.1051/0004-6361/201220307}{\JournalTitle{\aap},
  552, A49}

\bibitem[{{L\"{u}tzgendorf} {et~al.}(2013{\natexlab{b}}){L\"{u}tzgendorf},
  {Kissler-Patig}, {Neumayer}, {Baumgardt}, {Noyola}, {de Zeeuw}, {Gebhardt},
  {Jalali}, \& {Feldmeier}}]{lue13}
{L\"{u}tzgendorf}, N., {Kissler-Patig}, M., {Neumayer}, N., {et~al.}
  2013{\natexlab{b}},
  \href{http://dx.doi.org/10.1051/0004-6361/201321183}{\JournalTitle{\aap},
  555, A26}

\bibitem[{{MacLeod} {et~al.}(2016){MacLeod}, {Trenti}, \&
  {Ramirez-Ruiz}}]{mac16}
{MacLeod}, M., {Trenti}, M., \& {Ramirez-Ruiz}, E. 2016,
  \href{http://dx.doi.org/10.3847/0004-637X/819/1/70}{\JournalTitle{\apj}, 819,
  70}

\bibitem[{{Madau} \& {Rees}(2001)}]{mad01}
{Madau}, P., \& {Rees}, M.~J. 2001,
  \href{http://dx.doi.org/10.1086/319848}{\JournalTitle{\apjl}, 551, L27}

\bibitem[{{Madrid} {et~al.}(2017){Madrid}, {Leigh}, {Hurley}, \&
  {Giersz}}]{mad17}
{Madrid}, J.~P., {Leigh}, N.~W.~C., {Hurley}, J.~R., \& {Giersz}, M. 2017,
  \href{http://dx.doi.org/10.1093/mnras/stx1350}{\JournalTitle{\mnras}, 470,
  1729}

\bibitem[{{Mandel} {et~al.}(2008){Mandel}, {Brown}, {Gair}, \&
  {Miller}}]{man08}
{Mandel}, I., {Brown}, D.~A., {Gair}, J.~R., \& {Miller}, M.~C. 2008,
  \href{http://dx.doi.org/10.1086/588246}{\JournalTitle{\apj}, 681, 1431}

\bibitem[{{McKernan} {et~al.}(2014){McKernan}, {Ford}, {Kocsis}, {Lyra}, \&
  {Winter}}]{McKernan+2014}
{McKernan}, B., {Ford}, K.~E.~S., {Kocsis}, B., {Lyra}, W., \& {Winter}, L.~M.
  2014, \href{http://dx.doi.org/10.1093/mnras/stu553}{\JournalTitle{\mnras},
  441, 900}

\bibitem[{{McKernan} {et~al.}(2012){McKernan}, {Ford}, {Lyra}, \&
  {Perets}}]{McKernan+2012}
{McKernan}, B., {Ford}, K.~E.~S., {Lyra}, W., \& {Perets}, H.~B. 2012,
  \href{http://dx.doi.org/10.1111/j.1365-2966.2012.21486.x}{\JournalTitle{\mnras},
  425, 460}

\bibitem[{{Merritt}(2013)}]{mer13}
{Merritt}, D. 2013, {Dynamics and Evolution of Galactic Nuclei}

\bibitem[{{Merritt} \& {Ferrarese}(2001)}]{mer01}
{Merritt}, D., \& {Ferrarese}, L. 2001,
  \href{http://dx.doi.org/10.1086/318372}{\JournalTitle{\apj}, 547, 140}

\bibitem[{{Mezcua}(2017)}]{mez17}
{Mezcua}, M. 2017,
  \href{http://dx.doi.org/10.1142/S021827181730021X}{\JournalTitle{International
  Journal of Modern Physics D}, 26, 1730021}

\bibitem[{{Miller}(2002{\natexlab{a}})}]{mll02}
{Miller}, M.~C. 2002{\natexlab{a}},
  \href{http://dx.doi.org/10.1086/344156}{\JournalTitle{\apj}, 581, 438}

\bibitem[{{Miller}(2002{\natexlab{b}})}]{mil02a}
---. 2002{\natexlab{b}},
  \href{http://dx.doi.org/10.1086/344156}{\JournalTitle{\apj}, 581, 438}

\bibitem[{{Miller} \& {Hamilton}(2002)}]{mil02b}
{Miller}, M.~C., \& {Hamilton}, D.~P. 2002,
  \href{http://dx.doi.org/10.1046/j.1365-8711.2002.05112.x}{\JournalTitle{\mnras},
  330, 232}

\bibitem[{{Miller} \& {Miller}(2015)}]{mil15}
{Miller}, M.~C., \& {Miller}, J.~M. 2015,
  \href{http://dx.doi.org/10.1016/j.physrep.2014.09.003}{\JournalTitle{\physrep},
  548, 1}

\bibitem[{{Milosavljevi{\'c}} \& {Merritt}(2001)}]{mim01}
{Milosavljevi{\'c}}, M., \& {Merritt}, D. 2001,
  \href{http://dx.doi.org/10.1086/323830}{\JournalTitle{\apj}, 563, 34}

\bibitem[{{Moore} {et~al.}(2015){Moore}, {Cole}, \& {Berry}}]{moo15}
{Moore}, C.~J., {Cole}, R.~H., \& {Berry}, C.~P.~L. 2015,
  \href{http://dx.doi.org/10.1088/0264-9381/32/1/015014}{\JournalTitle{Classical
  and Quantum Gravity}, 32, 015014}

\bibitem[{{Morscher} {et~al.}(2015){Morscher}, {Pattabiraman}, {Rodriguez},
  {Rasio}, \& {Umbreit}}]{mor15}
{Morscher}, M., {Pattabiraman}, B., {Rodriguez}, C., {Rasio}, F.~A., \&
  {Umbreit}, S. 2015, \JournalTitle{\apj}, 800, 9

\bibitem[{{Morscher} {et~al.}(2013){Morscher}, {Umbreit}, {Farr}, \&
  {Rasio}}]{mor13}
{Morscher}, M., {Umbreit}, S., {Farr}, W.~M., \& {Rasio}, F.~A. 2013,
  \JournalTitle{\apj Lett.}, 763, 15

\bibitem[{{Navarro} {et~al.}(1997){Navarro}, {Frenk}, \& {White}}]{nfw97}
{Navarro}, J.~F., {Frenk}, C.~S., \& {White}, S.~D.~M. 1997,
  \href{http://dx.doi.org/10.1086/304888}{\JournalTitle{\apj}, 490, 493}

\bibitem[{{O'Leary} {et~al.}(2009){O'Leary}, {Kocsis}, \& {Loeb}}]{ole09}
{O'Leary}, R.~M., {Kocsis}, B., \& {Loeb}, A. 2009,
  \href{http://dx.doi.org/10.1111/j.1365-2966.2009.14653.x}{\JournalTitle{\mnras},
  395, 2127}

\bibitem[{{O'Leary} {et~al.}(2016){O'Leary}, {Meiron}, \& {Kocsis}}]{ole16}
{O'Leary}, R.~M., {Meiron}, Y., \& {Kocsis}, B. 2016,
  \href{http://dx.doi.org/10.3847/2041-8205/824/1/L12}{\JournalTitle{\apjl},
  824, L12}

\bibitem[{{O'Leary} {et~al.}(2006){O'Leary}, {Rasio}, {Fregeau}, {Ivanova}, \&
  {O'Shaughnessy}}]{ole06}
{O'Leary}, R.~M., {Rasio}, F.~A., {Fregeau}, J.~M., {Ivanova}, N., \&
  {O'Shaughnessy}, R. 2006,
  \href{http://dx.doi.org/10.1086/498446}{\JournalTitle{\apj}, 637, 937}

\bibitem[{{Pasham} {et~al.}(2014){Pasham}, {Strohmayer}, \&
  {Mushotzky}}]{pas14}
{Pasham}, D.~R., {Strohmayer}, T.~E., \& {Mushotzky}, R.~F. 2014,
  \href{http://dx.doi.org/10.1038/nature13710}{\JournalTitle{\nat}, 513, 74}

\bibitem[{{Peters}(1964)}]{pet64}
{Peters}, P.~C. 1964,
  \href{http://dx.doi.org/10.1103/PhysRev.136.B1224}{\JournalTitle{Physical
  Review}, 136, 1224}

\bibitem[{{Petts} \& {Gualandris}(2017)}]{pet17}
{Petts}, J.~A., \& {Gualandris}, A. 2017,
  \href{http://dx.doi.org/10.1093/mnras/stx296}{\JournalTitle{\mnras}, 467,
  3775}

\bibitem[{{Planck Collaboration}(2016)}]{planck15}
{Planck Collaboration}. 2016,
  \href{http://dx.doi.org/10.1051/0004-6361/201525830}{\JournalTitle{\aap},
  594, A13}

\bibitem[{{Portegies Zwart}(2006)}]{por05}
{Portegies Zwart}, S. 2006, {The Ecology of Black Holes in Star Clusters}, ed.
  F.~{Haardt}, V.~{Gorini}, U.~{Moschella}, \& M.~{Colpi}, 387

\bibitem[{{Portegies Zwart} \& {McMillan}(2000)}]{por00}
{Portegies Zwart}, S.~F., \& {McMillan}, S.~L.~W. 2000,
  \href{http://dx.doi.org/10.1086/312422}{\JournalTitle{\apjl}, 528, L17}

\bibitem[{{Portegies Zwart} \& {McMillan}(2002)}]{por02}
---. 2002, \href{http://dx.doi.org/10.1086/341798}{\JournalTitle{\apj}, 576,
  899}

\bibitem[{{Prieto} \& {Gnedin}(2008)}]{pri08}
{Prieto}, J.~L., \& {Gnedin}, O.~Y. 2008,
  \href{http://dx.doi.org/10.1086/591777}{\JournalTitle{\apj}, 689, 919}

\bibitem[{{Rodriguez} {et~al.}(2015){Rodriguez}, {Morscher}, {Pattabiraman},
  {Chatterjee}, {Haster}, \& {Rasio}}]{rod15}
{Rodriguez}, C.~L., {Morscher}, M., {Pattabiraman}, B., {et~al.} 2015,
  \href{http://dx.doi.org/10.1103/PhysRevLett.115.051101}{\JournalTitle{Physical
  Review Letters}, 115, 051101}

\bibitem[{{S{\'e}rsic}(1963)}]{ser63}
{S{\'e}rsic}, J.~L. 1963, \JournalTitle{Boletin de la Asociacion Argentina de
  Astronomia La Plata Argentina}, 6, 41

\bibitem[{{Sesana} {et~al.}(2012){Sesana}, {Sartore}, {Devecchi}, \&
  {Possenti}}]{ses12}
{Sesana}, A., {Sartore}, N., {Devecchi}, B., \& {Possenti}, A. 2012,
  \href{http://dx.doi.org/10.1111/j.1365-2966.2012.21958.x}{\JournalTitle{\mnras},
  427, 502}

\bibitem[{{Shapiro}(2005)}]{sha05}
{Shapiro}, S.~L. 2005,
  \href{http://dx.doi.org/10.1086/427065}{\JournalTitle{\apj}, 620, 59}

\bibitem[{{Sigurdsson} \& {Phinney}(1993)}]{sig93}
{Sigurdsson}, S., \& {Phinney}, E.~S. 1993,
  \href{http://dx.doi.org/10.1086/173190}{\JournalTitle{\apj}, 415, 631}

\bibitem[{{Sopuerta} {et~al.}(2007){Sopuerta}, {Yunes}, \& {Laguna}}]{sop07}
{Sopuerta}, C.~F., {Yunes}, N., \& {Laguna}, P. 2007,
  \href{http://dx.doi.org/10.1086/512067}{\JournalTitle{\apj Lett.}, 656, 9}

\bibitem[{{Spitzer}(1969)}]{spi69}
{Spitzer}, Jr., L. 1969,
  \href{http://dx.doi.org/10.1086/180451}{\JournalTitle{\apjl}, 158, L139}

\bibitem[{{Terzi{\'c}} \& {Graham}(2005)}]{ter05}
{Terzi{\'c}}, B., \& {Graham}, A.~W. 2005,
  \href{http://dx.doi.org/10.1111/j.1365-2966.2005.09269.x}{\JournalTitle{\mnras},
  362, 197}

\bibitem[{{Tiongco} {et~al.}(2016){Tiongco}, {Vesperini}, \& {Varri}}]{tio16}
{Tiongco}, M.~A., {Vesperini}, E., \& {Varri}, A.~L. 2016,
  \href{http://dx.doi.org/10.1093/mnras/stw1341}{\JournalTitle{\mnras}, 461,
  402}

\bibitem[{{Webb} {et~al.}(2014){Webb}, {Leigh}, {Sills}, {Harris}, \&
  {Hurley}}]{web14}
{Webb}, J.~J., {Leigh}, N., {Sills}, A., {Harris}, W.~E., \& {Hurley}, J.~R.
  2014, \href{http://dx.doi.org/10.1093/mnras/stu961}{\JournalTitle{\mnras},
  442, 1569}

\bibitem[{{Whalen} \& {Fryer}(2012)}]{wha12}
{Whalen}, D.~J., \& {Fryer}, C.~L. 2012,
  \href{http://dx.doi.org/10.1088/2041-8205/756/1/L19}{\JournalTitle{\apjl},
  756, L19}

\bibitem[{{Will}(2004)}]{wil04}
{Will}, C.~M. 2004,
  \href{http://dx.doi.org/10.1086/422387}{\JournalTitle{\apj}, 611, 1080}

\bibitem[{{Woods} {et~al.}(2017){Woods}, {Heger}, {Whalen}, {Haemmerl{\'e}}, \&
  {Klessen}}]{woo17}
{Woods}, T.~E., {Heger}, A., {Whalen}, D.~J., {Haemmerl{\'e}}, L., \&
  {Klessen}, R.~S. 2017,
  \href{http://dx.doi.org/10.3847/2041-8213/aa7412}{\JournalTitle{\apjl}, 842,
  L6}

\bibitem[{{Zaldarriaga} {et~al.}(2017){Zaldarriaga}, {Kushnir}, \&
  {Kollmeier}}]{zal17}
{Zaldarriaga}, M., {Kushnir}, D., \& {Kollmeier}, J.~A. 2017,
  \JournalTitle{ArXiv e-prints},
  \href{http://arxiv.org/abs/1702.00885}{{\sffamily arXiv:1702.00885
  [astro-ph.HE]}}

\end{thebibliography}

\end{document}